\documentclass{LMCS}

\def\doi{8 (2:02) 2012}
\lmcsheading%
{\doi}
{1--36}
{}
{}
{Nov.~15, 2009}
{Apr.~20, 2012}
{}

\usepackage{frameit}
\usepackage{epsfig}
\usepackage{latexsym}
\usepackage{wrapfig}
\usepackage{amsmath}
\usepackage{amssymb}
\usepackage{mathtools}
\usepackage{proof}
\usepackage{graphics}
\usepackage [geometry]{ifsym}
\usepackage{enumerate,hyperref}
\usepackage[nocenter]{qtree}

\newcommand{\hide}[1]{}
\newcommand{\myit}[1]{\textrm{#1}}

\newcommand{\code}[1]{{{\ensuremath{\tt #1}}}}
\newcommand{\mynew}[2]{#1\code{:=}~\code{new}~#2}
\newcommand{\barcall}[1]{{\bf \code{\bf barrier}}~#1}
\newcommand{\unit}{\texttt{skip}}
\newcommand{\myif}[3]{\code{if}~#1~\code{then}~#2~\code{else}~#3}
\newcommand{\mywhile}[2]{\code{while}~#1~\{#2\}}
\newcommand{\myfree}[1]{\code{free}~#1}

\newcommand{\myload}[2]{#1~\code{:=}~\code{[}#2\code{]}}
\newcommand{\myagn}[2]{#1~\code{:=}~#2}
\newcommand{\mystore}[2]{\code{[}#1\code{]}~\code{:=}~#2}
\newcommand{\myseq}[2]{#1\code{;}~ #2}
\def\lhalfp{\raisebox{-1pt}{\rotatebox[origin=c]{135}{\resizebox{3mm}{3mm}{\RightDiamond}}}}
\def\rhalfp{\raisebox{-1pt}{\rotatebox[origin=c]{315}{\resizebox{3mm}{3mm}{\RightDiamond}}}}

\def\fullp{\raisebox{0.5pt}{\rotatebox[origin=c]{90}{\resizebox{2.1mm}{2.1mm}{\FilledSquare}}}}
\def\emptyp{ \rotatebox[origin=c]{90}{\resizebox{3mm}{3mm}{\square}}}

\newcommand{\heapn}[3]{\ensuremath{#1 {\xmapsto{#2}} #3}}

\newcommand{\hoare}[3]{\ensuremath{\Gamma\vdash\{#1\}~#2~\{#3\}}}

\newcommand{\ptr}{\ensuremath{\gamma}}

\newcommand{\emp}{\btt{emp}}

\def\Vdash {\vdash}

\def\square{\rotatebox[origin=c]{0}{\SmallSquare}}
\def\sep{\code{*}}
\def\sept{~\code{*}~}
\def\constr{\ensuremath{\Phi}}
\def\pconstr{\ensuremath{\phi}}
\def\true{\code{true}}

\def\emp{{\mathsf{emp}}}

\def\heap{\ensuremath{\kappa}}
\def\nil{\mathsf{null}}
\def\D{\Delta}
\def\N{\mathbb{N}}

\newcommand{\view}[2]{\ensuremath{#1{\langle}{#2}{\rangle}}}
\newcommand{\barassert}[3]{\mathsf{barrier}(\mathit{#1},#2,\mathit{#3})}

\newcommand{\hdata}[3]{\mathit{#1}::#2{\langle}\mathit{#3}{\rangle}}
\newcommand{\hpred}[3]{#1::#2{\langle}\mathit{#3}{\rangle}}
\newcommand{\hpredhead}[2]{#1{\langle}\mathit{#2}{\rangle}}
\newcommand{\hfdata}[4]{\ensuremath{\mathit{#1}::#2^{#3}{\langle}\mathit{#4}{\rangle}}}
\newcommand{\hfpred}[4]{\ensuremath{#1::#2^{#3}{\langle}\mathit{#4}{\rangle}}}
\newcommand{\entailH}[3]{\ensuremath{#1{\vdash}#2\,{\sep}\,#3}}

\newcommand{\pure}{\ensuremath{\pi}}
\newcommand\term[1]{\mathit{#1}}
\def\ltree { \ensuremath{ \Tree [ $\bullet$ $\circ$ ] } }
\def\rtree { \ensuremath{ \Tree [ $\circ$ $\bullet$ ] } }

\newcommand{\entailK}[5]{\ensuremath{#3{\vdash}^{#1}_{#2}#4\,{\sep}\,#5}}
\newcommand{\entailVV}[3]{\entailK{\heap}{V}{#1}{#2}{#3}}

\newcommand{\sm}[1]{\mbox{$#1$}}
\newcommand{\btt}[1]{{\tt #1}}
\newcommand{\afrac}[2]{\frac{\begin{array}{c}#1\end{array}}{\begin{array}{c}#2\end{array}}}
\def\xpure {\mathsf{XPure}}
\def\fxpure {\mathsf{FXPure}}
\begin{document}

\title{\hspace{-1mm}Barriers in Concurrent Separation Logic:\\
 Now With Tool Support!}
\author{Aquinas Hobor}
\thanks{Aquinas Hobor is supported by a Lee Kuan Yew
Fellowship and MOE AcRF Tier 1 \#T1 251RES0902.}
\email {\{hobor, cristian\}@comp.nus.edu.sg}

\author{Cristian Gherghina}
\address{National University of Singapore}

\begin{abstract}
We develop and prove sound a concurrent separation logic for Pthreads-style barriers. Although
Pthreads barriers are widely used in systems, and separation logic is widely used for verification,
there has not been any effort to combine the two.  Unlike locks and critical sections, Pthreads
barriers enable simultaneous resource redistribution between multiple threads and are inherently
stateful, leading to significant complications in the design of the logic and its soundness proof.
We show how our logic can be applied to a specific example program in a modular way. Our proofs are
machine-checked in Coq. We showcase a program verification toolset that automatically applies the
logic rules and discharges the associated proof obligations.
\end{abstract}


\maketitle

\section {Introduction}

In a shared-memory concurrent program, threads communicate via a common memory. Programmers use
synchronization mechanisms, such as critical sections and locks, to avoid data races.  In a data
race, threads ``step on each others' toes'' by using the shared memory in an unsafe manner.
Recently, concurrent separation logic has been used to formally reason about shared-memory programs
that use critical sections and (first-class) locks
\cite{OHearn05resconc,HoborAN08,GotsmanBCRS07,Hobor:Thesis}.  Programs verified with concurrent
separation logic are provably data-race free.

What about shared-memory programs that use other kinds of synchronization mechanisms, such as
semaphores?  The general assumption is that other mechanisms can be implemented with locks, and
that reasonable Hoare rules can be derived by verifying their implementation. Indeed, the first
published example of concurrent separation logic was implementing semaphores using critical
sections \cite{OHearn05resconc}.  Unfortunately, not all synchronization mechanisms can be easily
reduced to locks in a way that allows for a reasonable Hoare rule to be derived.
In this paper  we introduce a Hoare rule that natively handles one such synchronization mechanism,
the Pthreads-style barrier.


Pthreads (POSIX Threads) is a widely-used API for concurrent programming, and includes various
procedures for thread creation/destruction and synchronization \cite{PThreadsBook}.  When a thread
issues a barrier call it waits until a specified number (typically all) of other threads have also
issued a barrier call; at that point, all of the threads continue.  Although barriers do not get
much attention in theory-oriented literature, they are very common in numerical applications code.
PARSEC is the standard benchmarking suite for multicore architectures, and has thirteen workloads
selected to provide a realistic cross-section for how concurrency is used in practice today; a
total of five (38\%) of PARSEC's workloads use barriers, covering the application domains of
financial analysis (blackscholes), computer vision (bodytrack), engineering (canneal), animation
(fluidanimate), and data mining (streamcluster) \cite{bienia10:phd}.  A common use for barriers is
to manage large numbers of threads in a pipeline setting.  For example, in a video-processing
algorithm, each thread might read from some shared common area containing the most recently
completed frame while writing to some private area that will contain some fraction of the next
frame.  (A thread might need to know what is happening in other areas of the previous frame to
properly handle objects entering or exiting its part of the current frame.) In the next iteration,
the old private areas become the new shared common area as the algorithm continues.

Our key insight is that a barrier is used to simultaneously redistribute ownership of resources
(typically, permission to read/write memory cells) between multiple threads.  In the
video-processing example, each thread starts out with read-only access to the previous frame and
write access to a portion of the current frame.  At the barrier call, each thread gives up its
write access to its portion of the (just-finished) frame, and receives back read-only access to the
entire frame.  Separation logic (when combined with fractional permissions
\cite{bornat05:popl,dockins09:sa}) can elegantly model this kind of resource redistribution. Let
$\mathit{Pre}_i$ be the preconditions that held upon entering the barrier, and $\mathit{Post}_i$ be
the postconditions that will hold after being released; then the following equation is
\emph{almost} true:
\begin{equation}
\label{badbarsep}
\raisebox{-7pt}{$\stackrel{\mbox{\huge \sep}}{_i}$} ~ {\vspace{-10pt}\mathit{Pre}_i} \qquad = \qquad
\raisebox{-7pt}{$\stackrel{\mbox{\huge \sep}}{_i}$} ~ {\vspace{-10pt}\mathit{Post}_i}
\end{equation}
Pipelined algorithms often operate in stages. Since barriers are used to ensure that one
computation has finished before the next can start, the barriers need to have stages as well---a
piece of ghost state associated with the barrier. We model this by building a finite automaton into
the barrier definition. We then need an assertion, written $\barassert{bn}{\pi}{cs}$, which says
that barrier $\mathit{bn}$, owned with fractional permission $\pi$, is currently in state
$\mathit{cs}$. The state of a barrier changes exactly as the threads are released from the barrier.
We can correct equation \eqref{badbarsep} by noting that barrier $\mathit{bn}$ is transitioning
from state $\mathit{cs}$ (current state) to state $\mathit{ns}$ (next state), and that the other
resources (frame $F$) are not modified:
\begin{equation}
\label{goodbarsep}
\vspace{-7pt} \quad ~
\begin{array}{lcl}
\raisebox{-7pt}{$\stackrel{\mbox{\huge \sep}}{_i}$} ~ {\vspace{-10pt}\mathit{Pre}_i} & \qquad = \qquad & F \sept \barassert{bn}{\fullp}{cs} \\
[5pt]
\raisebox{-7pt}{$\stackrel{\mbox{\huge \sep}}{_i}$} ~ {\vspace{-10pt}\mathit{Post}_i} & = & F \sept \barassert{bn}{\fullp}{ns}
\end{array}
\end{equation}
We use the symbol $\fullp$ to denote the full ($\sim$100\%) permission, which we require so that no
thread has a ``stale'' view of the barrier state. Although the on-chip (or \emph{erased})
operational behavior of a barrier is conceptually simple\footnote{Suspend each thread as it
arrives; keep a counter of the number of arrived threads; and when all of the threads have arrived,
resume the suspended threads.}, it may be already apparent that the verification can rapidly become
quite complicated.

\paragraph{Contributions.}
\begin{enumerate}[(1)]
  \item We give a formal characterization for sound barrier definitions.
  \item We design a natural Hoare rule in separation logic for verifying barrier calls.
  \item We give a formal resource-aware \emph{unerased} concurrent operational semantics for
  barriers and prove our Hoare rules sound with respect to our semantics.
  \item Our soundness results are machine-checked in Coq and are available at:
		\[{\texttt{www.comp.nus.edu.sg/{$\sim$}hobor/barrier}}\]
  \item We extended a program verification toolchain to automatically apply our Hoare rules to
      concurrent programs using barrier synchronization and discharge the resulting proof
      obligations. Our prototype is available at:
  		\[{\texttt{www.comp.nus.edu.sg/{$\sim$}cristian/projects/barriers/tool.html}}\]
\end{enumerate}

\paragraph{Relation to Previously Published Work.} We previously published on the design of the program
logic and its soundness proof \cite{hobor11:esop}; in \S\ref{sec:tools} this presentation additionally
presents our work on the modifications to the HIP/SLEEK program verifier we developed to reason about
our logic.

\hide{
			 \textbf{Contribution 1}: we give a formal
			characterization for sound barrier definitions.
			
			Although the barrier definitions can be nontrivial, one of the benefits of our approach is that the
			Hoare rules themselves are quite natural.  \textbf{Contribution~2}: we give an axiomatic semantics
			for barriers in concurrent separation logic.
			
			The on-chip (or \emph{erased}) operational behavior of a barrier is conceptually simple:
			suspend each thread as it arrives; keep a counter of the number of threads that have hit the
			barrier; and when all of the threads have arrived, resume the suspended threads. However, such a
			semantics does not allow for easy tracking of resource permissions.
\textbf{Contribution 3}: we
			define a formal resource-aware concurrent (\emph{unerased}) operational semantics for barriers.
			
			We connect the axiomatic semantics to our operational semantics by using the oracle semantics of
			Hobor \emph{et al.}.  \textbf{Contribution 4}: we prove our Hoare rules sound with respect to our
			operational semantics.  The soundness proof for the Hoare rule relating to the barrier instruction
			is complex. \textbf{Contribution 5}: our soundness results are machine-checked in Coq.  Our proofs
			are available at:
			\[\texttt{www.comp.nus.edu.sg/{$\sim$}hobor/barrier}\]
}

\section{Syntax, Separation Algebras, Shares, and Assertions}
\label{sec:synsharsep}

Here we briefly introduce preliminaries: the syntax of our language, separation algebras, share
accounting, and the assertions of our separation logic.
\subsection{Programming Language Syntax}

To let us focus on the barriers, most of our programming language is pure vanilla.  We define four
kinds of (tagged) values $v$: \texttt{TRUE}, \texttt{FALSE}, \texttt{ADDR}($\N$), and
\texttt{DATA}($\N$). We have two (tagged) expressions $e$: \texttt{C}($v$) and \texttt{V}($x$),
where $x$ are local variable names (just $\N$ in Coq).  To make the example more interesting we add
the arithmetical operations to $e$.  We write $\mathsf{bn}$ for a barrier number, with $\mathsf{bn}
\in \N$.

We have ten commands $c$: {\unit} (do nothing), \myagn{$x$}{$e$} (local variable assignment),
\myload{$x$}{$e$} (load from memory), \mystore{$e_1$}{$e_2$} (store to memory), \mynew{$x$}{$e$}
(memory allocation), \myfree{$e$} (memory deallocation), \myseq{$c_1$}{$c_2$} (instruction
sequence), \myif{$e$}{$c_1$}{$c_2$} (if-then-else), \mywhile{$e$}{$c$} (loops), and
\barcall{$\mathsf{bn}$} (wait for barrier $\mathsf{bn}$).  To run commands $c_1$ \ldots $c_n$ in
parallel (which, like O'Hearn, we only allow at the top level \cite{OHearn05resconc}), we write
$c_1 || \ldots || c_n$.  To avoid clogging the presentation, we elide a setup sequence before the
parallel composition.

\subsection{Disjoint Multi-unit Separation Algebras}

Separation algebras are mathematical structures used to model separation logic.  We use a variant
described by Dockins \emph{et al}. called a disjoint multi-unit separation algebra (hereafter just
``DSA'') \cite{dockins09:sa}.  Briefly, a DSA is a set $S$ and an associated three-place partial
\emph{join relation} $\oplus$, written $x \oplus y = z$, such that:

\begin{tabular}{lcl}
A function: && $x \oplus y = z_1 ~~ \Rightarrow ~~ x \oplus y = z_2 ~~ \Rightarrow ~~ z_1 = z_2$ \\
Commutative:&& $x \oplus y ~~ = ~~ y \oplus x$ \\
Associative:&& $x \oplus (y \oplus z) ~~ = ~~ (x \oplus y) \oplus z$ \\
Cancellative:&& $x_1 \oplus y = z ~~ \Rightarrow ~~ x_2 \oplus y = z ~~ \Rightarrow ~~ x_1 = x_2$ \\
Multiple units: &\qquad& $\forall x.~ \exists u_x.~ x \oplus u_x = x$ \\
Disjointness: && $x \oplus x = y ~~ \Rightarrow ~~ x = y$ \\
\end{tabular}

\noindent A key concept is the idea of an \emph{identity}: $x$ is an identity if  $x \oplus y =
z$ implies $y = z$.  One fundamental property of identities is that $x$ is an identity if and
only if $x \oplus x = x$.  Dockins also develops a series of standard constructions (\emph{e.g.}, product,
functions, etc.) for building complicated DSAs from simpler DSAs.  We make use of this idea to
construct a variety of separation algebras as needed, usually with the concept of \emph{share} as
the ``foundational'' DSA.

\subsection{Shares} \label{sec:fracperm}

Separation logic is a logic of \emph{resource ownership}.  Concurrent algorithms sometimes want to
have threads share some common resources.  Bornat \emph{et al}. introduced the concept of
\emph{fractional share} to handle the necessary accounting \cite{bornat05:popl}.  Shares form a
DSA; a \emph{full share} (complete ownership of a resource) can be broken into various
\emph{partial shares}; these shares can then be rejoined into the full share. The \emph{empty
share} is the identity for shares.  We often need non-empty (strictly \emph{positive}) shares,
denoted by $\pi$. A critical invariant is that the sum of each thread's share of a given object is no more or
less than the full share.

The semantic meaning of partial shares varies; here we use them in two distinct ways.   We require
the full share to modify a memory location; in contrast, we only require a positive share to read
from one.  There is no danger of a data race even though we do not require the full share to read:
if a thread has a positive share of some location, no other thread can have a full share for the
same location. We use fractional permissions differently for barriers: each precondition includes
some positive share of the barrier itself and we require that the preconditions combine to imply
the full share of the barrier (plus a frame $F$).

In the Coq development we use a share model developed Dockins \emph{et al}. that supports
sophisticated fractional ownership schemes \cite{dockins09:sa}.  Here we simplify this model into
four elements: the full share {\fullp}; two \textbf{distinct} nonempty partial shares, {\lhalfp}
and {\rhalfp}, and the empty share {\emptyp}.  The key point is that $\lhalfp \oplus \rhalfp =
\fullp$.


\subsection{Assertion Language}\label{sec:ass_lang}

We model the assertions of separation logic following Dockins \emph{et al}. \cite{dockins09:sa}.
Our states $\sigma$ are triples of a store, heap, and barrier map ($\sigma = (s,h,b)$). Local
variables live in stores $s$ (functions from variable names to values).  In contrast, a heap $h$
contains the locations shared between threads; heaps are partial functions from addresses to pairs
of positive shares and values.  We also equip our heaps with a distinguished location, called the
\emph{break}, that tracks the boundary between allocated and unallocated locations. The break lets
us provide semantics for the \mynew{$x$}{$e$} instruction in a natural way by setting $x$ equal to
the current break and then incrementing the break.  Since threads share a common break, there is a
covert communication channel (one thread can observe when another thread is allocating memory);
however the existence of this channel is a small price to pay for avoiding the necessity of a
concurrent garbage collector. We ensure that the threads see the same break by equipping our break
with ownership shares just as we equip normal memory locations with shares.

We denote the empty heap (which lacks ownership for both all memory locations and the distinguished
break location) by $h_0$. Of note, our expressions $e$ are evaluated only in the context of the
store; we write $s \vdash e \Downarrow v$ to mean that $e$ evaluates to $v$ in the context of the
store $s$. Finally, the barrier map $b$ is a partial function from barrier numbers to pairs of
barrier states (represented as natural numbers) and positive shares; we denote the empty barrier
map by $b_0$.

An \emph{assertion} is a function from states to truth values (\texttt{Prop} in Coq).  As is
common, we define the usual logical connectives via a straightforward embedding into the metalogic;
for example, the object-level conjunction $P \wedge Q$ is defined as $\lambda \sigma.~ (P \sigma)
\wedge (Q \sigma)$.  We will adopt the convention of using the same symbol for both the
object-level operators and the meta-level operators to avoid symbol bloat; it should be clear from
the context which operator applies in a given situation.  We provide all of the standard
connectives ($\top,\bot,\wedge,\vee,\Rightarrow,\neg,\forall,\exists$).

We model the connectives of separation logic in the standard way\footnote{Our Coq definition for
$\mathsf{emp}$ is different but equivalent to the definition given here.}:
\[
\begin{array}{lcl}
\mathsf{emp}  &=& \lambda (s,h,b).~ h=h_0 ~ \wedge ~ b=b_0 \\ 
P \sept Q  &=& \lambda \sigma.~ \exists \sigma_1, \sigma_2. ~ \sigma_1 \oplus \sigma_2 = \sigma ~ \wedge ~ P(\sigma_1) ~ \wedge ~ Q(\sigma_2) \\
e_1 \xmapsto{\pi} e_2 &=& \lambda (s, h, b).~ \exists a,v. ~ (s \vdash e_1\Downarrow \texttt{ADDR}(a)) ~ \wedge ~ (s \vdash e_2\Downarrow v)~ \wedge ~ \\
&& ~ b = b_0 ~ \wedge ~ h(a) = (v, \pi) ~ \wedge ~ \mathsf{dom}(h) = \{a\} ~ \wedge ~ \mathsf{break}(h)=\emptyp \\
\barassert{\mathit{bn}}{\pi}{s} &~ =~ &\lambda (s,h,b).~ h = h_0 ~ \wedge ~ b(\mathit{bn}) = (s, \pi) ~ \wedge ~ \mathsf{dom}(b) = \{\mathit{bn}\}
\end{array}
\]
The fractional points-to assertion, $e_1 \xmapsto{\pi} e_2$, means that the expression $e_1$ is
pointing to an address $a$ in memory; $a$ is owned with positive share $\pi$, and contains the
evaluated value $v$ of $e_2$. The fractional points-to assertion does not include any ownership of
the break. The barrier assertion, $\barassert{\mathit{bn}}{\pi}{s}$, means that the barrier
$\mathit{bn}$, owned with positive share $\pi$, is in state $s$.

We also lift program expressions into the logic: $e \hspace{-1pt} \Downarrow \hspace{-1pt} v$,
which evaluates $e$ with $\sigma$'s store (\emph{i.e.}, $\lambda (s, h, b).~ h = h_0 \wedge b = b_0
\wedge s \vdash e \hspace{-1pt} \Downarrow \hspace{-1pt} v$);  $[e]$, equivalent to $e
\hspace{-1pt} \Downarrow \hspace{-1pt} \texttt{TRUE}$; and $x=v$, equivalent to $\texttt{V}(x)
\hspace{-1pt} \Downarrow \hspace{-1pt} v$.  These assertions have a ``built-in'' $\mathsf{emp}$.


\section {Example}\label{sec:example}

\begin{figure}[!!h]
\begin{center}
\begin{minipage}{13cm}
\begin{center}
\begin{tabular}{l l@{\extracolsep{\fill}}  || l@{\extracolsep{\fill}}  }
0:& \multicolumn{2}{c} {$\hspace{-15mm}\{\heapn{x_1}{\fullp}{0}\sept\heapn{x_2}{\fullp}{0}
 \sept\heapn{y_1}{\fullp}{0} \sept\heapn{y_2}{\fullp}{0} \sept
 \heapn{i}{\fullp}{0} \sept \barassert{bn}{\fullp}{0} \} $ }\\
[5pt]
0':~~&$\{\heapn{x_1}{\lhalfp}{0}\sept\heapn{x_2}{\lhalfp}{0}\sept\heapn{y_1}{\lhalfp}{0}$ ~~ & ~~
$\{\heapn{x_1}{\rhalfp}{0}\sept\heapn{x_2}{\rhalfp}{0}\sept\heapn{y_1}{\rhalfp}{0}$ \\
&\hspace*{-20pt} $\sept\heapn{y_2}{\lhalfp}{0} \sept \heapn{i}{\lhalfp}{0} \sept
\barassert{bn}{\lhalfp}{0}\}~ $ & $\sept\heapn{y_2}{\rhalfp}{0} \sept
\heapn{i}{\rhalfp}{0} \sept \barassert{bn}{\rhalfp}{0}\} $\\
[3pt] ~~& \ldots  & ~~ \ldots \\
1:~~& \barcall{b}; & ~~\barcall{b}; ~~~~~~~~// b transitions 0$\rightarrow$1\\
2:~~& \myagn{n}{0}; & ~~\myagn{m}{0};\\
3:~~& \code{while} $ n<30$ \{ & ~~\code{while} $ m<30$ \{\\
4:~~&  $~~~~~ \myload{a_1}{x_1}; ~~~$ & $~~~~~~~\myload{a_1}{x_1};$ \\
5:~~&  $~~~~~ \myload{a_2}{x_2}; ~~~$ & $~~~~~~~\myload{a_2}{x_2};$ \\
6:~~&  $~~~~~ \mystore{y_1}{~(a_1{+}2{*}a_2)}; ~~~$ & $~~~~~~~\mystore{y_2}{~(a_1{+}3{*}a_2)};$ \\
7:~~&  ~~~~~\barcall{b}; & ~~~~~~~\barcall{b}; ~~~// b transitions 1$\rightarrow$2\\
8:~~& $~~~~~ \myload{a_1}{y_1}; ~~~$ & $~~~~~~~\myload{a_1}{y_1};$ \\
9:~~& $~~~~~ \myload{a_2}{y_2}; ~~~$ & $~~~~~~~\myload{a_2}{y_2};$ \\
10:~~& $~~~~~ \mystore{x_1}{~(a_1{+} 2{*}a_2)}; ~~~$ & $~~~~~~~\mystore{x_2}{~(a_1{+}3{*}a_2)};$ \\
11:~~& $~~~~~ \myagn{n}{(n{+}1)};$ &  \\
12:~~& $~~~~~ \mystore{i}{n};$ & \\
13:~~& ~~~~~\barcall{b}; & ~~~~~~~\barcall{b}; ~~// b transitions 2$\rightarrow$1\\
14:~~& & ~~~~~~~\myload{m}{i};\\
15:~~& \}&~~\}\\
16:~~& \barcall{b}; & ~~ \barcall{b}; ~~~~~~~~~// b transitions 1$\rightarrow$3\\
17:~~& \mystore{i}{0}; \\
  ~~& \ldots  & ~~ \ldots \\
  \\
\end{tabular}
\vspace{-5mm}
{\centering
\includegraphics*[viewport=0 0 1300 920,scale=.27]{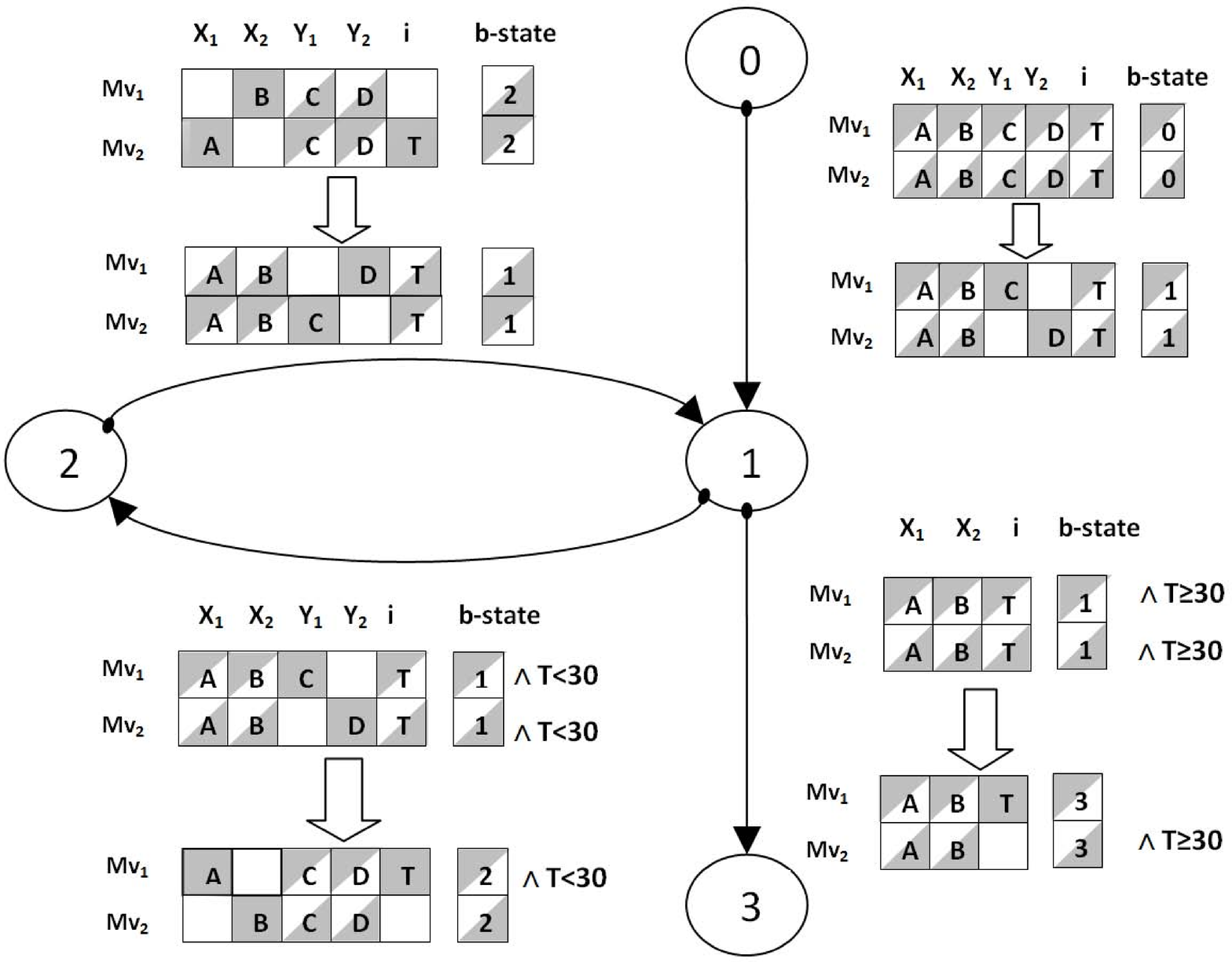}
}
\end{center}
\end{minipage}
\end{center}
\caption{Example: Code and Barrier Diagram} \label{fig:ex_code}\label{fig:state_img}
\vspace{-3mm}
\end{figure}

We present a detailed example inspired by a video decompression algorithm. The code and a
detailed-but-informal description of the barrier definition is given in Figure
\ref{fig:ex_code}.\footnote{In our Coq development we give the full formal description of the
example barrier.} Two threads cooperate to repeatedly compute the elements of two size-two arrays
\code{x} and \code{y}. In each iteration, each thread writes to a single cell of the ``current''
array, and reads from both cells of the ``previous'' array.

In Figure \ref{fig:state_img} we give a pictorial representation of the state machine associated
with the barrier used in the code using the following specialized notation:
\[
\includegraphics*[viewport=10 0 1185 350,scale=.28]{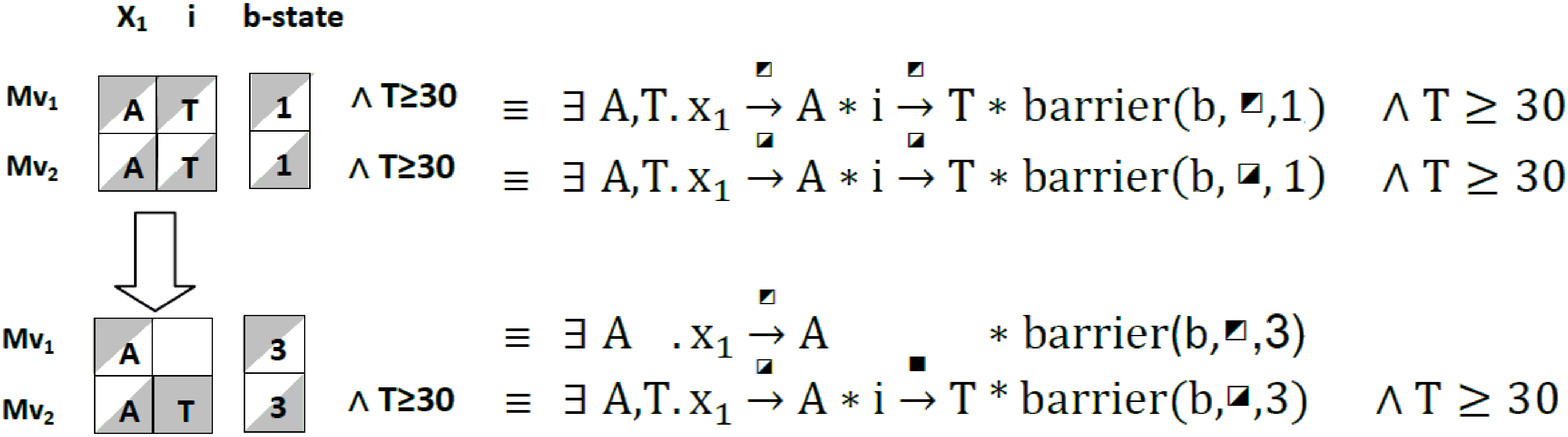}
\]
This notation is used to express the pre- and postconditions for a given barrier transition.  Each
row is a pictorial representation (values, barrier states, and shares) of a formula in separation
logic as indicated above.  The preconditions are on top (one per row) and the postconditions below.
Each row is associated with a \emph{move}; move 1 is a pair of the first precondition row and the
first postcondition row, etc. A barrier that is waiting for $n$ threads will have $n$ moves; $n$
can be fewer than the total number of threads.  We do not require that a given thread always takes
the same move each time it reaches a given barrier transition.

Note that only the permissions on the memory cells change during a transition; the contents
(values) do not.\footnote{We use the same quantified variable names before and after the transition
because an outside observer can tell that the values are the same.  A local verification can use
ghost state to prove the equality; alternatively we could add the ability to move the quantifier to
other parts of the diagram, \emph{e.g.}, over an entire pre-post pair.} The exception to this is
the special column on the right side, which denotes the assertion associated with the barrier
itself. As the barrier transitions, this value changes from the previous state to the next; we
require that the sum of the preconditions includes the full share of the barrier assertion to
guarantee that no thread has an out-of-date view of the barrier's state. Observe that all of the
preconditions join together, and, except for the state of the barrier itself, are exactly equal to
the join of the postconditions.

The initial state of the machine is given as an assertion in line 0. The machine starts with full
ownership of the array cells $x_1$, $x_2$, $y_1$, and $y_2$, as well as an additional cell $i$,
used as a condition variable.  The barrier b is fully-owned and is in state 0. The initial state is
then partitioned into two parts on line 0', with the left thread (A) and right thread (B) getting
the shares {\lhalfp} and {\rhalfp}, respectively.



Not shown (between lines 0' and 1) is thread-specific initialization code; perhaps both threads
read both arrays and perform consistency checks.  The real action starts with the barrier call on
line 1, which ensures that this initialization code has completed.  Thread A takes move 1 and
thread B takes move 2.  Afterwards, thread A has full ownership over $y_1$ and thread B has full
ownership over $y_2$; the ownership of $x_1$, $x_2$, and $i$ remains split between A and B.  While
the ownership of the barrier is unchanged, it is now in state 1.

We then enter the main loop on line 3.  On lines 4--5, both threads read from the shared cells
$x_1$ and $x_2$, and on line 6 both threads update their fully-owned cell. The barrier call on line
7 ensures that these updates have been completed before the threads continue.  Since the value T at
memory location $i$ is less than 30, only the 1--2 transition is possible; the 1--3 transition
requires T$\ge 30$.  Thread A takes move 1 and thread B takes move 2\footnote{In this example a
given thread always takes the same move for a given transition; however, this is not forced by the
rules of our logic.}; afterwards, both threads have partial shares of $y_1$ and $y_2$, thread A has
the full share of $x_1$ and the condition cell $i$, and thread B has the full share of $x_2$; the
barrier is in state 2.

Lines 8--10 are mirrors of lines 4--6.  On lines 11--12, thread A updates the condition cell $i$.
The barrier on line 13 ensures that the updates on lines 10 and 12 have completed before the
threads continue; thread A takes move 2 while thread B takes move 1.  Afterwards, the threads have
the same permissions they had on entering the loop: A has full ownership of $y_1$, B has full
ownership of $y_2$, and they share ownership of $x_1$, $x_2$, and $i$; the barrier is again in
state 1.

On line 14, thread B reads from the condition variable $i$, and then the program loops back to line
3.  After 30 iterations, the loop exits and control moves to the barrier on line 16.  Observe that
since the (shared) value T at memory location $i$ is greater than or equal to 30, only the 1--3
transition is possible; the 1--2 transition requires T$<30$.  Thread A takes move 1 while thread B
takes move 2; afterwards, both threads are sharing ownership of $x_1$, $x_2$, $y_1$, and $y_2$
(since the transition from 1 to 3 does not mention $y_1$ and $y_2$ they are unchanged). Thread A
has full permission over the condition variable $i$; the barrier is in state 3.  Finally, on line
17, thread A updates $i$; the barrier on line 16 ensures that thread B's read of $i$ on line 14 has
already occurred.

\section {Barrier Definitions and Consistency Requirements} \label{sec:bardefs}


We present the type of a barrier definition in Figure \ref{fig:bar_def} in the form of a data
structure.  The definitions include numerous consistency requirements; in Coq these are maintained
with dependent types. From the top down, a barrier definition ($\mathsf{BarDef}$) consists of a
barrier identifier (\emph{i.e.}, barrier number), the number of threads the barrier is
synchronizing, and a list of barrier state definitions.  For programs that have more than one
barrier, the individual barrier definitions will be collected into a list and barrier number $j$
will be in list slot $j$.

\begin{figure}[t]
\begin{center}
\begin{minipage}{13cm}
\begin{center}
\[
\begin{array}{lclrl}
\mathsf{BarDef} &\quad\equiv\quad & \{ ~\mathsf{bd\_bn}: \mathit{Nat} & \mathrm{barrier ~id}\\
\mathrm{(barrier~definition)}&& ~~\mathsf{bd\_limit} : \mathit{Nat} & \qquad \mathrm{\# ~of ~threads} & \qquad \\
&& ~~\mathsf{bd\_states} : \mathit{list}~\mathsf{BarStateDef}\} & \mathrm{state ~list}\\
\mathsf{BarStateDef} &\equiv& \{\mathsf{bsd\_bn} : \mathit{Nat} & \mathrm{barrier ~id}\\
\mathrm{(barrier ~state)}&& ~~\mathsf{bsd\_cs}: \mathit{Nat} & \mathrm{state ~id}\\
&&~~\mathsf{bsd\_directions} : \mathit{list}~\mathsf{BarMoveList} & \mathrm{transition ~list}\\
&&~~\mathsf{bsd\_limit} : \mathit{Nat}\}  &\mathrm{\# ~of ~threads}\\
\mathsf{BarMoveList} &\equiv& \{\mathsf{bml\_ns} : \mathit{Nat} & \mathrm{next ~state}\\
\mathrm{(transition)} &&~~\mathsf{bml\_bn} : \mathit{Nat} & \mathrm{barrier ~id}\\
&&~~\mathsf{bml\_cs} : \mathit{Nat} & \mathrm{current ~state}\\
&&~~\mathsf{bml\_limit} : \mathit{Nat} &\mathrm{\# ~of ~threads}\\
&&~~\mathsf{bml\_moves} : \mathit{list}~(\mathsf{assert} \times \mathsf{assert}) \} & \mathrm{pre/post ~pairs}\\
 \end{array}
\]
\end{center}
\end{minipage}
\caption{Barrier Definitions}
\label{fig:bar_def}
\end{center}
\end{figure}

A barrier state definition ($\mathsf{BarStateDef}$) consists of a barrier number, the number of
threads synchronized, a state id, and a transition list; such that:
\begin{enumerate}[(1)]
\item the barrier number matches the barrier number in the containing $\mathsf{BarDef}$
\item the limit matches the limit of the containing $\mathsf{BarDef}$\footnote{A command to
    dynamically alter the number of threads a barrier managed might allow different
    states/transitions to wait for different numbers of threads.}
\item the state identifier $j$ indicates that this $\mathsf{BarStateDef}$ is the $j$ element of
    the containing $\mathsf{BarDef}$'s list of state definitions
\item the directions are \emph{mutually exclusive}
\end{enumerate} 
 The first three are unexciting; we will discuss mutual exclusion shortly.

A transition ($\mathsf{BarMoveList}$) contains a barrier number ($\mathsf{bn}$), number of threads
synchronized, current state identifier ($\mathsf{cs}$), next state identifier ($\mathsf{ns}$), and
list of precondition/postcondition pairs (the \emph{move list}).  We require that:
\begin{enumerate}[(1)]
\item $\mathsf{bn}$ matches the barrier number in the containing $\mathsf{BarStateDef}$
\item the limit matches the limit in the containing $\mathsf{BarStateDef}$
\item $\mathsf{cs}$ matches the state identifier in the containing $\mathsf{BarStateDef}$
\item the length of list of moves ($\mathsf{bml\_moves}$) is equal to the limit
    ($\mathsf{bml\_limit}$)
\item all of the pre/postconditions in the movelist ignore the store, focusing only on the
    memory and barrier map. Since stores are private to each thread (on a processor these would
    be registers), it does not make sense for them to be mentioned in the ``public'' pre/post
    conditions.
\item all of the preconditions in the movelist are \emph{precise}.  Precision is a technical
    property involving the identifiability of states satisfying an assertion. An assertion $P$
    is precise when
    \[
    \infer
    {\sigma_1 = \sigma1'}
    {\sigma_1 \oplus \sigma_2 = \sigma_3 \qquad \sigma_1 \models P \qquad \sigma_1' \oplus \sigma_2' = \sigma_3 \qquad \sigma_1' \models P}
    \]
    That is, $P$ can hold on at most one substate of an arbitrary state $\sigma_3$.
    \footnote{Precision may not be required; another property (tentatively christened
    ``token'') that might serve would be if, for any precondition $P$, $P \sept P \equiv \bot$.
    Note that precision in conjunction with item (6) implies $P$ is a token.}
\item each precondition $P$ includes some positive share of the $\mathsf{barrier}$ assertion
    with $\mathsf{bn}$ and $\mathsf{cs}$, \emph{i.e.}, $\exists \pi.~ P \Rightarrow \top \sept
    \barassert{\mathsf{bn}}{\pi}{\mathsf{cs}}$.
\item the sum of the preconditions must equal the sum of the postconditions, except for the
    state of the barrier; moreover, the sum of the preconditions must include the full share of
    the barrier (equation \eqref{goodbarsep}, repeated here):
    \[
    \begin{array}{lcl}
    \raisebox{-7pt}{$\stackrel{\mbox{\huge \sep}}{_i}$} ~ {\vspace{-10pt}\mathit{Pre}_i} & \qquad = \qquad & F \sept \barassert{\mathsf{bn}}{\fullp}{\mathsf{cs}} \\
    [5pt]
    \raisebox{-7pt}{$\stackrel{\mbox{\huge \sep}}{_i}$} ~ {\vspace{-10pt}\mathit{Post}_i} & = & F \sept \barassert{\mathsf{bn}}{\fullp}{\mathsf{ns}}
    \end{array}
    \]
\end{enumerate} 
Items 1--4 are simple bookkeeping; items 5--7 are similar to technical requirements
required in other variants of concurrent separation logic
\cite{OHearn05resconc,Hobor:Thesis,GotsmanBCRS07}. As previously mentioned, the fundamental insight
of this approach is property (8).


The function $\mathsf{lookup\_move}$ simplifies the lookup of a move in a $\mathsf{BarDef}$:
\[
\mathsf{lookup\_move}(\mathit{bd},\mathit{cs},\mathit{dir},\mathit{mv}) = bd.\mathsf{bd\_states}[\mathit{cs}].\mathsf{bsd\_directions}[\mathit{dir}].\mathsf{bml\_moves}[\mathit{mv}]
\]
Using this notation, we can express the important requirement that all directions in the barrier
state $\mathit{cs}$ of the barrier definition $\mathsf{bd}$ are mutually exclusive:
\[
\begin{array}{l}
\forall \mathit{dir}_1, \mathit{dir}_2, \mathit{mv}_1, \mathit{mv}_2, \mathit{pre}_1, \mathit{pre}_2.~ \mathit{dir}_1 \neq \mathit{dir}_2 ~ \Rightarrow \\
~~ \mathsf{lookup\_move}(\mathit{bd},\mathit{cs},\mathit{dir}_1,\mathit{mv}_1) = (\mathit{pre}_1, \underline{~~}) ~ \Rightarrow \\
~~~~ \mathsf{lookup\_move}(\mathit{bd},\mathit{cs},\mathit{dir}_2,\mathit{mv}_2) = (\mathit{pre}_2, \underline{~~}) ~ \Rightarrow \\
[3pt]
~~~~~~~~ (\top \sept \mathit{pre}_1) ~ \wedge ~ (\top \sept \mathit{pre}_2) ~~ \equiv ~~ \bot
\end{array}
\]
In other words, it is \emph{impossible} for any of the preconditions of more than one transition
(of a given state) to be true at a time.  The simplest way to understand this is to consider the
1--2 and 1--3 transitions in the example program.  The 1--2 transition requires that the value in
memory cell $i$ be strictly less than 30; in contrast, the 1--3 transition requires that \emph{the
same cell} contains a value greater than or equal to 30.  Plainly these are incompatible; but in
fact the above property is stronger: \emph{both} of the moves on the 1--2 transition, and
\emph{both} of the moves on the 1--3 transition include the incompatibility.  Thus, if thread A
takes transition 1--2, it knows for certain that thread B \emph{cannot} take transition 1--3.  This
way we ensure that both threads always agree on the barrier's current state.

\section{Hoare Logic}\label{sec:hoare}

Our Hoare judgment has the form $\hoare{P}{c}{Q}$, where $\Gamma$ is a list of barrier definitions
as given in \S \ref{sec:bardefs}, $P$ and $Q$ are assertions in separation logic, and $c$ is a
command.  Our Hoare rules come in three groups: standard Hoare logic (Skip, If, Sequence, While,
Assignment, Consequence); standard separation logic (Frame, Store, Load, New, Free); and the
barrier rule.  We give all three groups two and three in Figure \ref{fig:hoare_seq}. We note four
points for group two.

\begin{figure}[t]
\begin{center}
\begin{minipage}{13cm}
\begin{center}
\[
\begin{array}{c}

\infer[\mathrm{Skip}]{\hoare{P}{\unit}{P}}{}  \qquad
 \infer[\mathrm{If}]{\hoare{P}{\myif{e}{c_t}{c_f}}{Q}}{\hoare{P \sept
 [e]}{c_t}{Q} \quad \hoare{P \sept \neg [e]}{c_f}{Q}}\\
 \\
 \\
 \infer[\mathrm{Seq.}]{\hoare {P} {c_1 ~; ~c_2} {R}}{\hoare{P} {c_1} {Q}
\quad \hoare{Q}{c_2}{R}}
\quad
 \infer[\mathrm{While}]{\hoare{I}{\mywhile{e}{c}}{I \sept \neg [e]}}{\hoare{I \sept [e]}{c}{I}} \\
\\
 \infer[\mathrm{Assign}]{\hoare {e \hspace{-2pt} \Downarrow \hspace{-2pt} v } {\myagn{x}{e}} {x \hspace{-1pt}= \hspace{-1pt} v}}{} \quad
 \infer[\mathrm{Conseq.}]{\hoare{P'}{c}{Q'}}{P' \vdash P ~~ \hoare{P}{c}{Q} ~~ Q \vdash Q'} \\
\\

  \infer[\mathrm{Frame}]{\hoare{F \sept P} {c} {F\sept Q}}{\hoare{P}{c}{Q} ~~ \mathsf{closed} (F,c)} ~~~~
  \infer[\mathrm{Store}]{\hoare{\heapn{e_1}{\fullp}{\_}}{\mystore{e_1}{e_2}}{\heapn{e_1}{\fullp}{e_2}}}{~}\\
\\
  \infer[\mathrm{Load}]{
  \hoare{\heapn{e_1}{\pi}{e_2} \sept e_1 \hspace{-2pt}\Downarrow \hspace{-2pt} v_1 \sept e_2 \hspace{-2pt} \Downarrow \hspace{-2pt} v_2}
  {\myload{x}{e_1}}
  {\heapn{\texttt{C}(v_1)}{\pi}{\texttt{C}(v_2)} \sept x = v_2}
  }{}\\
\\
  \infer[\mathrm{New}]{\hoare{e \hspace{-2pt} \Downarrow \hspace{-2pt} v}{\mynew{x}{e}}{\heapn{\texttt{V}(x)}{\fullp}{\texttt{C}(v)}}}{} ~~
  \infer[\mathrm{Free}]{\hoare{\heapn{e_1}{\fullp}{e_2}}{\myfree{e_1}}{\mathsf{emp}}}{}\\
\\
\fbox{
 \infer[\mathrm{Barrier}]{\hoare{\mathit{P}}{\barcall{bn}}{\mathit{Q}}}
 {\Gamma[\mathit{bn}] = \mathit{bd} \qquad \mathsf{lookup\_move}(\mathit{bd},\mathit{cs},\mathit{dir},\mathit{mv}) = (P,Q)}
 }
\end{array}
\]
\end{center}
\end{minipage}
\caption{Hoare rules}
\label{fig:hoare_seq}
\end{center}
\end{figure}

First, as explained in \S \ref{sec:ass_lang}, the assertions $e \Downarrow v$, $[e]$ and $x = v$
are bundled with an assertion that the heap and barrier map are empty(\emph{i.e.}, $e \Downarrow v
~~ \Rightarrow ~~ \mathsf{emp}$); thus, we use the separating conjunction when employing them.
Second, the rules are in ``side-condition-free form''.  Thus, instead of presenting the load rule
as $\hoare{\heapn{e_1}{\pi}{e_2}}{\myload{x}{e_1}}{x = e_2 \sept \heapn{e_1}{\pi}{e_2}}$, which is
aesthetically attractive but untrue in the pesky case when $e_2$ depends on $x$ (\emph{e.g.},
$\myload{x}{x}$), we use a form that is less visually pleasing but does not require side
conditions.\footnote{Recall from \S \ref{sec:synsharsep}: $\texttt{V}(x)$ and $\texttt{C}(v)$ are
expression constructors for locals and constants.  In addition, $\mathsf{closed} (F,c)$ means that
$F$ does not depend on locals modified by $c$.} It is straightforward to restore rules with side
conditions via the Consequence rule. Third, our Store and Free rules require the full share of
location $e_1$; in contrast, our Load rule only requires some positive share; this is consistent
with our use of fractional permissions as explained in \S \ref{sec:fracperm}.  Fourth, memory
allocation and deallocation are more complicated in concurrent settings than in sequential
settings, and so the New and Free rules cause nontrivial complications in the semantic model.




The Hoare rule for barriers is so simple that at first glance it may be hard to understand.  The
variables for the current state $\mathit{cs}$, direction $\mathit{dir}$, and move $\mathit{mv}$
appear to be free in the $\mathsf{lookup\_move}$!  However, things are not quite as unconstrained
as they initially appear.  Recall from \S \ref{sec:bardefs} that one of the consistency
requirements for the precondition $P$ is that $P$ implies an assertion about the barrier itself: $P
\Rightarrow Q \sept \barassert{\mathit{bn}}{\pi}{\mathit{cs}}$; thus at a given program point we
can only use directions and moves from the current state.  Similarly, recall from \S
\ref{sec:bardefs} that since the directions are mutually exclusive, $\mathit{dir}$ is uniquely
determined.

This leaves the question of the uniqueness of $\mathit{mv}$.  If a thread only satisfies a single
precondition, then the move $\mathit{mv}$ is uniquely determined.  Unfortunately, it is simple to
construct programs in which a thread enters a barrier while satisfying the preconditions of
multiple moves.  What saves us is that we are developing a logic of partial correctness.  Since
preconditions to moves must be precise and nonempty (\emph{i.e.}, token), only one thread is able
to satisfy a given precondition at a time.  The pigeonhole principle guarantees that if a thread
holds multiple preconditions then some other thread will not be able to enter the barrier; in this
case, the barrier call will never return and we can guarantee any postcondition.








We now apply the Barrier rule to the barrier calls in line 13 from our example program; the
$\mathsf{lookup\_move}$s are direct from the barrier state diagram:
\[
\begin{array}{l}
\mathsf{Thread ~A} ~
\begin{cases}
\infer{\hoare{P}{\barcall{b}}{Q}}
{\begin{array}{c}
 \mathsf{lookup\_move}(b,2,1,2) = (P,Q)\\
 P = \heapn{y_1} {\lhalfp} {v_{y1}} \sept \heapn{y_2} {\lhalfp} {v_{y2}} \sept \heapn {x_1}
 		{\fullp} {v_{x1}} \sept \heapn{i} {\fullp} {v_i} \sep \barassert{bn}{\lhalfp}{2} \\
 Q = \heapn{y_1} {\fullp} {v_{y1}} \sept \heapn{x_1} {\lhalfp} {v_{x1}} \sept \heapn {x_2} {\lhalfp}
		{v_{x2}} \sept \heapn{i} {\lhalfp} {v_i} \sep \barassert{bn}{\lhalfp}{1}\\
 \end{array} }
\end{cases}\\
\mathsf{Thread ~B} ~
\begin{cases}
	\infer{\hoare{P}{\barcall{b}}{Q}}
	{
	\begin{array}{c}
		\mathsf{lookup\_move}(b,2,1,1) = (P,Q)\\
		P = \heapn {y_1} {\rhalfp} {v_{y1}} \sept \heapn {y_2} {\rhalfp} {v_{y2}} \sept \heapn {x_2}
			{\fullp} {v_{x2}} \sep\barassert{bn}{\rhalfp}{2}\}\vspace{3pt}\\
		Q = \heapn {y_2}{\fullp} {v_{y2}} \sept \heapn {x_1} {\rhalfp} {v_{x1}} \sept \heapn {x_2}
			{\rhalfp} {v_{x2}} \sept \heapn{i} {\rhalfp} {v_i} \sep \barassert{bn}{\rhalfp}{1}\}\\
	\end{array}
	}
\end{cases}\\
\end{array}
\]
Note that in this line of the example program, the frame is $\mathsf{emp}$ in both threads.


Not shown in Figure \ref{fig:hoare_seq} is a parallel composition rule. As in \cite{Hobor:Thesis},
each thread is verified independently using the Hoare rules given; a top-level safety theorem
proves that the entire concurrent machine behaves as expected.

\section {Semantic Models}\label{sec:semantic}

Our operational semantics is divided into three parts: purely sequential, which executes all of the
instructions except for barrier in a thread-local manner; concurrent, which manages thread
scheduling and handles the barrier instruction; and oracular, which provides a pseudosequential
view of the concurrent machine to enable simple proofs of the sequential Hoare rules.  Our setup
follows Hobor \emph{et al}. very closely and we refer readers there for more detail
\cite{HoborAN08,Hobor:Thesis}.


\paragraph{Purely sequential semantics.}  The purely sequential semantics executes the
instructions \unit, \myagn{$x$}{$e$}, \myload{$x$}{$e$}, \mystore{$e_1$}{$e_2$}, \mynew{$x$}{$e$},
\myfree{$e$}, \myseq{$c_1$}{$c_2$}, \texttt{if} $e$ \texttt{then} $c_1$ \texttt{else} $c_2$, and
\mywhile{$e$}{$c$}. The form of the sequential step judgment is $(\sigma, c) \mapsto (\sigma',c')$.
Here $\sigma$ is a state (triple of store, heap, barrier map), just as in \S \ref{sec:ass_lang} and
$c$ is a command of our language.  The semantics of the sequential instructions is standard; the
only ``tricky'' part is that the machine gets stuck if one tries to write to a location for which
one does not have full permission or read from a location for which one has no permission;
\emph{e.g.}, here is the store rule:
\[
\infer[\mathsf{sstep-store}]
{\big( (s,h,b),~ \myseq{\mystore{e_1}{e_2}}{c}\big) ~ \mapsto ~ \big((s,h',b),c \big)}
{
\begin{array}{c}
s \vdash e_1 \Downarrow \texttt{C}(\texttt{ADDR}(n)) \qquad s \vdash e_2 \Downarrow v \\
n < \mathsf{break}(h) \qquad h(n) = (\fullp,v') \qquad h' = [n \mapsto (\fullp,v)]h
\end{array}
}
\]
The test that $n < \mathsf{break}(h)$ ensures that the address for the store is ``in
bounds''---that is, less than the current value of the break between allocated and unallocated
memory; since we are updating the memory we require that the permission associated with the
location $n$ be full ($\fullp$).  We say that this step relation is \emph{unerased} since these
bounds and permission checks are virtual rather than on-chip.


We define the other cases of the step relation in a similar way. Observe that if we were in a
sequential setting the proof of the Hoare store rule would be straightforward; this is likewise the
case for the other cases of the sequential step relation and their associated Hoare rules.  If the
sequential step relation reaches a barrier call \barcall{$\mathsf{bn}$} then it simply gets stuck.


\paragraph{Concurrent semantics.}  We define the notion of a \emph{concurrent state} in Figure
\ref{fig:cstate}.  A concurrent state contains a scheduler $\Omega$ (modeled as a list of natural
numbers), a distinguished heap called the \emph{allocation pool}, a list of \emph{thread}s, and a
\emph{barrier pool}\footnote{There is also a series of consistency requirements such as the fact
that all of the heaps in the threads and barrier pool join together with the allocation pool into
one consistent heap; in the mechanization this is carried around via a dependent type as a fifth
component of the concurrent state.  We elide this proof from the presentation.}. The allocation
pool ``owns'' all of the unallocated memory cells and the ``break'' that indicates the division
between allocated and unallocated cells.  Before we run a thread we transfer the allocation pool
into the local heap owned by the thread so that \texttt{new} can transfer a cell from this pool
into the local heap of a thread when required. When we suspend the thread we remove (what is left
of) the allocation pool from its heap so that we can transfer it to the next thread.

A thread contains a (sequential) state (store, heap, and barrier map) and a \emph{concurrent
control}, which is either $\mathsf{Running}(c)$, meaning the thread is available to run command
$c$, or $\mathsf{Waiting} (\mathit{bn}, \mathit{dir}, \mathit{mv},c)$, meaning that the thread is
currently waiting on barrier $\mathit{bn}$ to make move $\mathit{mv}$ in direction $\mathit{dir}$;
after the barrier call completes the thread will resume running with command $c$.

\begin{figure}[t]
\begin{center}
\hspace{-8.5mm}\begin{minipage}{13cm}
\begin{center}
\[
\begin{array}{lclr}
\mathsf{Cstate} &\quad\equiv\quad& \{ ~\mathsf{cs\_sched}: \mathsf{list ~\N} &
\mathrm{schedule}\\
&&~~\mathsf{cs\_allocpool}: \mathsf{heap} & \mathrm{alloc ~pool}\\
&&~~\mathsf{cs\_thds}: \mathsf{list}~\mathsf{Thread}  & \mathrm{thread ~pool}\\
&&~~\mathsf{cs\_barpool}: \mathsf{Barpool}\} & \mathrm{barrier ~pool}\\
[3pt]\mathsf{Thread} &\quad\equiv\quad& \{ ~\mathsf{th\_stk}: \mathsf{store} &\\
&&~~\mathsf{th\_hp}: \mathsf{heap} &\\
&&~~\mathsf{th\_bs}: \mathsf{BarrierMap} & \mathrm{local~view ~of~barrier ~states}\\
&&~~\mathsf{th\_ctl}: \mathsf{conc\_ctl}\}& \mathrm{running ~or ~waiting}\\
[3pt]\mathsf{conc\_ctl} &\quad\equiv\quad&
   | \mathsf{~Running}(c) & \mathrm{executing ~code ~c}\\
&& | \mathsf{~Waiting}(bn,dir,mv,c)  & \mathrm{waiting~on~bn}\\
[3pt]\mathsf{Barpool} &\quad\equiv\quad&\{~\mathsf{bp\_bars}: \mathsf{list}~\mathsf{DyBarStatus}
&\quad\mathrm{ dynamic~barrier~status} \\
 &&~~\mathsf{bp\_st}: \mathsf{store \times heap \times BarrierMap}\} & \mathrm{current ~state}\\
[3pt]\mathsf{DyBarStatus}&\quad\equiv\quad& \{~\mathsf{dbs\_bn}: \mathit{\N}& \mathrm{barrier~id}\\
&&~~\mathsf{dbs\_wp}: \mathsf{Waitpool}&\mathrm{ waiting~thread~pool}\\
&&~~\mathsf{dbs\_bd}: \mathsf{BarDef}\}& \\
[3pt]\mathsf{Waitpool}&\quad\equiv\quad& \{~\mathsf{wp\_dir}: \mathit{option ~ \N}&
\mathrm{direction ~id}\\
&& ~~\mathsf{wp\_slots}: \mathit{option~}(\mathsf{list~} \mathsf{slot})& \mathrm{taken ~slots}\\
&& ~~\mathsf{wp\_limit}: \mathit{\N}&\\
&& ~~\mathsf{wp\_st}: \mathsf{store \times heap \times BarrierMap}\} & \mathrm{current ~state}\\
[3pt]\mathsf{slot}&\quad\equiv\quad& \mathsf{(thread\_id \times heap \times BarrierMap)}
&\qquad\qquad\mathrm{ waiting~slot}\\
\end{array}
\]
\end{center}
\end{minipage}
\caption{Concurrent state} \label{fig:cstate}
\end{center}
\end{figure}

The barrier pool ($\mathsf{Barpool}$) contains a list of \emph{dynamic barrier status}es (DBSes) as
well as a state which is the join of all of the states inside the DBSes.  Each DBS consists of a
barrier number (which must be its index into the array of its containing $\mathsf{Barpool}$), a
barrier definition (from \S \ref{sec:bardefs}), and a \emph{waitpool} (WP).  A waitpool consists of
a direction option (\texttt{None} before the first barrier call in a given state; thereafter the
unique direction for the next state), a limit (the number of threads synchronized by the barrier,
and comes from the barrier definition in the enclosing DBS), a \emph{slot} list, and a state (which
is the join of all of the states in the slot list).  A slot is a heap and barrier map (the store is
unneeded since barrier pre/postconditions ignore it) as well as a thread id (whence the heap and
barrier map came as a precondition, and to which the postcondition will return).

The concurrent step relation \hide{judgment} has the form
$(\Omega, \mathit{ap}, \mathit{thds}, \mathit{bp}) \leadsto (\Omega', \mathit{ap}', \mathit{thds}',
\mathit{bp}')$,
where $\Omega$, $\mathit{ap}$, $\mathit{thds}$, and $\mathit{bp}$ are the scheduler, allocation
pool, thread list, and barrier pool respectively.  The concurrent step relation has only four
cases; the following case CStep-Seq is used to run all of the sequential commands:
\[
\infer[\hspace{-50pt}\raisebox{-20pt}{\textrm{CStep-Seq}}]
{(i::\Omega, \mathit{thds}, \mathit{ap}, \mathit{bp}) \leadsto (i::\Omega, \mathit{thds}', \mathit{ap}', \mathit{bp})}
{
\begin{array}{c}
\mathit{thds}[i] = (s,h,b,\mathsf{Running}(c)) \qquad h \oplus \mathit{ap} = h' \qquad
\big((s,h',b),c \big) \mapsto \big((s',h'',b), c' \big) \\
h''' \oplus \mathit{ap}' = h'' \qquad \mathsf{isAllocPool}(\mathit{ap}') \qquad
\mathit{thds}' = [i \mapsto (s', h''', b, \mathsf{Running}(c'))] \mathit{thds}
\end{array}
}
\]
That is, we look up the thread whose thread id is at the head of the scheduler, join in the
allocation pool, and run the sequential step relation.  If the command $c$ is a barrier call then
the sequential relation will not be able to run and so the $\textrm{CStep-Seq}$ relation will not
hold; otherwise the sequential step relation will be able to handle any command.  After we have
taken a sequential step, we subtract out the (possibly diminished) allocation pool, and reinsert
the modified sequential state into the thread list.  Since we quantify over all schedulers and our
language does not have input/output, it is sufficient to utilize a non-preemptive scheduler; for
further justification on the use of such schedulers see \cite{Hobor:Thesis}.

The second case of the concurrent step relation handles the case when a thread has reached the last
instruction, which must be a \unit:
\[
\infer[\textrm{CStep-Exit}]
{(i::\Omega, \mathit{thds}, \mathit{ap}, \mathit{bp}) \leadsto (\Omega, \mathit{thds}, \mathit{ap}, \mathit{bp})}
{\mathit{thds}[i] = \mathsf{Running}(\unit)}
\]
When we reach the end of a thread we simply context switch to the next thread.

The interesting cases occur when the instruction for the running thread is a barrier call; here the
CStep-Seq rule does not apply.  The concurrent semantics handles the barrier call directly via the
last two cases of the step relation; before presenting these cases we will first give a technical
definition called $\mathsf{fill\_barrier\_slot}$:
\[
\infer{\mathsf{fill\_barrier\_slot}  ~~(thds, ~bp, ~bn, ~i) = (thds',~bp')}{
\begin{array}{c}
  thds[i] = \mathsf{Thread}(stk, ~hp, ~bs, ~(\mathsf{Running} ~(\barcall{bn}  ; ~c)))\\
  [2pt]\mathsf{lookup\_move} (bp.bp\_bars[bn], ~dir, ~mv) = (pre,post)\\
  hp' \oplus hp'' = hp \qquad\quad bs' \oplus bs'' = bs \qquad\quad pre(stk,hp',bs')\\
  [2pt]\mathsf{bp\_inc\_waitpool} ~(bp, ~bn, ~dir, ~mv, ~(i, ~(hp', ~bs'))) = bp' \\
  [2pt]thds' = [i\rightarrow (\mathsf{Thread} (~stk ,~hp'', ~bs'', ~(\mathsf{Waiting} ~(bn, ~dir,
  ~mv, ~c))))]~thds\\
\end{array}
}
\]
The predicate $\mathsf{fill\_barrier\_slot}$ gives the details of removing the (sub)state
satisfying the precondition of the barrier from the thread's state, inserting it into the barrier
pool, and suspending the calling thread.  The predicate $\mathsf{bp\_inc\_waitpool}$ does the
insertion into the barrier pool; the details of manipulating the data structure are straightforward
but lengthy to formalize\footnote{In Coq things are trickier since we track some technical side
conditions via dependent types so this relation also ensures that these side conditions remain
satisfied.}.

We are now ready to give the first case for the barrier, used when a thread executes a barrier but
is not the last thread to do so:
\[
\infer[\textrm{CStep-Suspend}]{((i :: \Omega), ap, thds, bp) \leadsto (\Omega, ap, thds', bp')}
   {
     \begin{array}{c}     	
     	\mathsf{fill\_barrier\_slot}  ~~(thds, ~bp, ~bn, ~i) = (thds',~bp')\\
     	\neg ~\mathsf{bp\_ready} ~~(bp', ~bn)\\    	
     \end{array}}
\]
After using $\mathsf{fill\_barrier\_slot}$, CStep-Suspend checks to see if the barrier is full by
counting the number of slots that have been filled in the appropriate wait pool by using the
$\mathsf{bp\_ready}$ predicate, and then context switches.

If the barrier is ready then instead of using the CStep-Suspend case of the concurrent step
relation, we must use the CStep-Release case:
\[
\infer[\textrm{CStep-Release}]{((i :: \Omega), ap, thds, bp)\leadsto(\Omega, ap, thds'', bp'')}
		{
		\begin{array}{c}	    	
    	\mathsf{fill\_barrier\_slot}  ~~(thds, ~bp, ~bn, ~i) = (thds',~bp')\\
    	\mathsf{bp\_ready} ~~(bp', ~bn)\\    	
    	\mathsf{bp\_transition} ~~(bp', ~bn, ~out) = bp''\\
    	\mathsf{transition\_threads} ~~(out, ~thds') = thds''\\
	\end{array}}
\]
The first requirement of CStep-Release is exactly the same as CStep-Suspend: we suspend the thread
and transfer the appropriate resources to the barrier pool.  However, now all of the threads have
arrived at the barrier and so it is ready.  We use the $\mathsf{bp\_transition}$ predicate to go
through the barrier's slots in the $\mathsf{waitpool}$, combine the associated heaps and barrier
maps, redivide these resources according to the barrier postconditions, and remove the associated
resources from the barrier pool into a list of slots called $\mathsf{out}$.  Finally, the states in
$\mathsf{out}$ are combined with the suspended threads, which are simultaneously resumed by the
$\mathsf{transition\_threads}$ predicate.  The formal definitions of the $\mathsf{bp\_transition}$
and $\mathsf{transition\_threads}$ predicates are extremely complex and very tedious and we refer
interested readers to the mechanization.





\paragraph{Oracle semantics.}  Following Hobor \emph{et al.}
\cite{HoborAN08,Hobor:Thesis}, we define a third \emph{oracular semantics}: $( \sigma,o,c ) \mapsto
 ( \sigma',o',c' )$.
Here the sequential state $\sigma$ and command $c$ are exactly the same as in the purely sequential
step.  The new parameter $o$ is an oracle, a kind of box containing ``the rest'' of the concurrent
machine---that is, $o$ contains a scheduler, a list of other threads, and a barrier pool.

The oracle semantics behaves exactly the same way as the purely sequential semantics on all of the
instructions except for the barrier call, with the oracle $o$ being passed through unchanged.  That
is to say:
\[
\infer[\textrm{os-seq}]
{\big( \sigma,o,c \big) ~ \mapsto ~ \big( \sigma',o,c' \big)}
{\big( \sigma,c \big) ~ \mapsto ~ \big( \sigma',c' \big)}
\]
When the oracle semantics reaches a barrier instruction, it consults the oracle $o$ to determine
the state of the machine after the barrier:
\[
\infer[\textrm{os-consult}]
{\big( (s,h,b),o,\myseq{\barcall{\mathsf{bn}}}{c} \big) ~ \mapsto ~ \big( (s,h',b'),o',c \big)}
{\mathsf{consult}(h,b,o) = (h',b',o')}
\]
The formal definition of the $\mathsf{consult}$ relation is detailed in
\cite{HoborAN08,Hobor:Thesis} but the idea is simple.  To consult the oracle, one unpacks the
concurrent machine and runs (classically) all of the other threads until control returns to the
original thread; consult then returns the current $h'$ and $b'$ (that resulted from the barrier
call) and repackages the concurrent machine into the new oracle $o'$.  The final case of the oracle
semantics occurs when the concurrent machine never returns control (because it got stuck or due to
sheer perversion of the scheduler):
\[
\infer[\textrm{os-diverge}]
{\big( (s,h,b),o,\myseq{\barcall{\mathsf{bn}}}{c} \big) ~ \mapsto ~ \big( (s,h,b),o,\myseq{\barcall{\mathsf{bn}}}{c} \big)}
{\raisebox{1pt}{$\not$} \exists r.~ \mathsf{consult}(h,b,o) = r \qquad \textrm{(\emph{i.e.}, $\mathsf{consult}$ diverges)}}
\]
When control will never return, it does not matter what this thread does as long as it does not get
stuck; accordingly we enter an (infinite) loop.

\paragraph{Soundness proof outline.} Our soundness argument falls into several parts.  We define
our Hoare tuple in terms of our oracle semantics using a definition by Appel and Blazy
\cite{appel07:tphols}; this definition was designed for a sequential language and we believe that
other standard sequential definitions for Hoare tuples would work as well\footnote{We change Appel
and Blazy's definition so that our Hoare tuple guarantees that the allocation pool is available for
verifying the Hoare rule for \mynew{$x$}{$e$}.}. We then prove (in Coq) all of the Hoare rules for
the sequential instructions; since the os-seq case of the oracle semantics provides a straight lift
into the purely sequential semantics this is straightforward\footnote{The Hoare rule for loops
(While) is only proved on paper.  The loop rule is known to be painful to mechanize and so the
mechanization was skipped due to time constraints. It has been proved in Coq for similar (indeed,
more complicated from a sequential control-flow perspective) settings in previous work
\cite{appel07:tphols,HoborAN08}.}.

Next, we prove (in Coq) the soundness for the barrier rule.  This turns out to be much more
complicated than a proof of the soundness of (non-first-class) locks and took the bulk of the
effort.  There are two points of particular difficulty: first, the excruciatingly painful
accounting associated with tracking resources during the barrier call as they move from a source
thread (as a precondition), into the barrier pool, and redistribution to the target thread(s) as
postcondition(s).  The second difficulty is proving that a thread that enters a barrier while
holding more than one precondition will never wake up; the analogy is a door with $n$ keys
distributed among $n$ owners; if an owner has a second key in his pocket when he enters then one of
the remaining owners will not be able to get in.

After proving the Hoare rules from Figure \ref{fig:hoare_seq} sound with respect to the oracle
semantics, the remaining task is to connect the oracle semantics to the concurrent semantics---that
is, \emph{oracle soundness}. Oracle soundness says that if each of the threads on a machine are
safe with respect to the oracle semantics, then the entire concurrent machine combining the threads
together is safe. The (very rough) analogy to this result in Brookes' semantics is the parallel
decomposition lemma. Here we use a progress/preservation style proof closely following that given
in \cite[\emph{pp}.242--255]{Hobor:Thesis}; the proof was straightforward and quite short to
mechanize. A technical advance over previous work is that the progress/preservation proofs do not
require that the concurrent semantics be deterministic.  In fact, allowing the semantics to be
nondeterministic simplified the proofs significantly.

A direct consequence of oracle soundness is that if each thread is verified with the Hoare rules,
and is loaded onto a single concurrent machine, then if the machine does not get stuck and if it
halts then all of the postconditions hold.

\paragraph{Erasure.} One can justly observe that our concurrent semantics is not especially
realistic; \emph{e.g.}, we: explicitly track resource ownership permissions (\emph{i.e.}, our
semantics is \emph{unerased}); have an unrealistic memory allocator/deallocator and scheduler;
ignore issues of byte-addressable memory; do not store code in the heap; and so forth. We believe
that we could connect our semantics to a more realistic semantics that could handle each of these
issues, but most of them are orthogonal to barriers.  For brevity we will comment only on erasing
the resource accounting since it forms the heart of our soundness result.

We have defined, in Coq, an \emph{erased} sequential and concurrent semantics.  An erased memory is
simply a pair of a break address and a total function from addresses to values.  The run-time state
of an erased barrier is simply a pair of naturals: the first tracking the number of threads
currently waiting on the barrier, and the second giving the final number of threads the barrier is
waiting for.  We define a series of $\mathsf{erase}$ functions that take an unerased type
(memory/barrier status/thread/etc.) to an erased one by ``forgetting'' all permission information.
The sequential erased semantics is quite similar to the unerased one, with the exception that we do
not check if we have read/write permission before executing a load/store. The concurrent erased
semantics is much simpler than the complicated accounting-enabled semantics explained above since
all that is needed to handle the barrier is incrementing/resetting a counter, plus some modest
management of the thread list to suspend/resume threads.  Critically, our erased semantics is a
computable function, enabling program evaluation.  Finally, we have proved that our unerased
semantics is a conservative approximation to our erased one: that is, if our unerased concurrent
machine can take a step from some state $\Sigma$ to $\Sigma'$, then our erased machine takes a step
from $\mathsf{erase}(\Sigma)$ to $\mathsf{erase}(\Sigma')$.

 \hide{
			 Based on the above definitions let us now define when does a Hoare tuple hold. For this we
			 first describe a bit of machinery. Let $\tau$ be the state of one sequential machine, executing one
			 thread, $\tau = (s,h,b\_map)$. We define a configuration, $\sigma$ as a triple of a sequential
			 machine state, an oracle and the rest of the code to be executed by the current thread, $\sigma =
			 (\tau, o, k)$.
			
			\begin{definition}
				We define the $\mathsf{brk\_pred}$ predicate as\\
				 ~$\qquad\mathsf{brk\_pred}(brk) = \lambda \sigma. (brk = Some ~ m)\wedge \forall n.
				 (n>m \rightarrow ~h_\sigma(n)=Some(\_,0))$.
			\end{definition}
			
			 The $\mathsf{brk\_pred}$ predicate serves as the repository of the unused memory, any thing that
			is above the current memory watermark. It is from this repository that the allocator will choose a
			new location. This predicate will be required as a precondition for executing any fresh step. This
			is in accordance with the machine semantic, as described in \S\ref{sec:semantic}.
			
			 We define the $ora\_barDefs$ predicate that takes an oracle and a barrier definition list and
			 enforces a wellformed condition on the oracles. This ensures that the barrier definitions contained within
			 the oracle is consistent.
			
			\begin {definition}
				For any configuration $c=(\tau,o,k)$:
				\[
				\begin{array}{lll}
			    	\mathsf{can\_step} ~c &:=& \exists c'. ~ \mathsf{seq\_step} ~c ~c' \\
					\mathsf{safely\_halted} ~c &:=& ~(snd ~c ~=~ Unit) \\
					\mathsf{safe\_ora}(\tau, o, k) &:=& ~\forall ~\tau' ~o' ~k'. ~\mathsf{seq\_stepstar}
					(\tau,o,k)(\tau',o',k') {\rightarrow} \mathsf{can\_step} ~(\tau',o',k') \\
					&&\qquad\qquad\vee~ \mathsf{safely\_halted} ~(\tau',o',k')\\
					\mathsf{safe}(\Gamma, ~\tau, ~k) &:=& \forall o.  ~\mathsf{ora\_barDefs} ~(o,~\Gamma)
					\rightarrow \mathsf{safe\_ora} ~(\tau, ~o, ~k)
			    \end{array}
				\]
			\end{definition}
			
			We defined a safe configuration as a configuration in which the sequential machine is either safely
			halted, the oracle does not prescribe any further steps, or if there is a possible next step then
			the new configuration is a configuration that has at least one more successor, that is the machine
			does not get stuck.
				
			\begin{definition} We say that P guards k in context $\Gamma$ and write  $\Gamma \Vdash P\square k $
			if for all possible states that satisfy $P\sep \mathsf{brk\_pred}(brk)$ those states are safe with
			respect to command k and context $\Gamma$.\\
				$P\square k = \forall \tau. \tau \Vdash P \sep \mathsf{brk\_pred}(brk)  \Rightarrow
				\mathsf{safe}(\Gamma,\tau,k)$
			\end{definition}
			
			\begin{definition}\label{def:hoare}{\bf(Hoare Tuple)}
			 	We say that $\hoare{P}{k}{Q}$ holds if, given a context $\Gamma$,  for all frames F and commands
			 	k' following k, if $(F\sep Q)\square k'$ then $P \square k\cdot k'$.\\
			 	$\hoare{P}{k}{Q} \approx \forall k', F.  (F\sep Q)\square k' \Rightarrow F\sep P \square k\cdot
			 	k'$.
			\end{definition}
}

\section {Coq Development} \label{sec:coq}

\begin{figure}[t]
\begin{center}
\begin{minipage}{13cm}
\begin{center}
\[
\begin{tabular}{|l|c|c|c|}
\hline
File & LOC & Time & Description\\
\hline
\texttt{SLB\_Base} & ~~1,182~~ & ~~~~2s~~~~ & Utility lemmas (largely list facts) \\
\texttt{SLB\_Lang} & 1,240 & 11s & States, program syntax, assertion model \\
\texttt{SLB\_BarDefs} & 265 & 2s & Barrier definitions \\
\texttt{SLB\_CLang} & 3,230 & 1m7s & Dynamic concurrent state \\
\texttt{SLB\_SSem} & 415 & 17s & Sequential semantics \\
\texttt{SLB\_Sem} & 784 & 33s & Concurrent semantics \\
\texttt{SLB\_ESSem} & 230 & 5s & Erased semantics \\
\texttt{SLB\_ESEquiv} & 3,352 & 30s & Erasure proofs \\
\texttt{SLB\_OSem} & 1,942 & 2m10s & Oracular semantics \\
\texttt{SLB\_HRules} & 170 & 2s & Definition of Hoare tuples \\
\texttt{SLB\_OSound} & 426 & 30s & Soundness of oracle semantics \\
\texttt{SLB\_HRulesSound} ~~& 1,664 & 1m14s & Soundness proofs for Hoare rules \\
\texttt{SLB\_Ex} & 2,700 & 48s & Example of a barrier definition \\
\hline
Total & 16,598 & 7m34s & \\
\hline
\end{tabular}
\]
\end{center}
\end{minipage}
\caption{Proof structure, size and compilation times (2.66GHz, 8GB)}
\label{tbl:proof}
\end{center}
\end{figure}

We detail our Coq development in Figure \ref{tbl:proof}.  We use the Mechanized Semantic Library
\cite{MSL} for the definitions of share models, separation algebras, and various utility
lemmas/tactics.  In addition to the standard Coq axioms, we use dependent and propositional
extensionality and the law of excluded middle.

Over 7,000 lines of the development is devoted to proving the soundness of the Hoare rule for
barriers, largely in the files \texttt{SLB\_BarDefs.v}, \texttt{SLB\_CLang.v}, \texttt{SLB\_Sem.v},
\texttt{SLB\_OSem.v}, \texttt{SLB\_HRules.v}, and a small part of \texttt{SLB\_HRulesSound.v}. The
rest of the concurrent semantics, the oracle semantics, and the soundness of the oracle semantics
($\sim$the parallel decomposition lemma) require approximately 1,000 lines, largely in the files
\texttt{SLB\_Sem.v}, \texttt{SLB\_HRules.v}, and \texttt{SLB\_OSound}. The erased semantics
requires 230 lines in \texttt{SLB\_ESSem.v}, while the associated equivalence proofs
require 3352 lines in the file \texttt{SLB\_ESEquiv.v}.

The sequential semantics and proofs for the associated Hoare rules require approximately 2,000
lines drawn from the files \texttt{SLB\_Lang.v}, \texttt{SLB\_SSem.v}, \texttt{SLB\_HRules.v}, and
\texttt{SLB\_HRulesSound.v}.  We estimate that the proof of the loop rule would require a further
2,000-3,000 lines.  The model of our assertions and the program syntax are both in
\texttt{SLB\_Lang.v}. Utility lemmas/tactics (\texttt{SLB\_Base.v}) and the example barrier
(\texttt{SLB\_Ex.v}) complete the development.

\section{Tool support}
\label{sec:tools}

We have integrated our program logic for barriers into the HIP/SLEEK program verification toolset
\cite{ChinCav08,FM11:Gherghina}.  SLEEK is an entailment checker for separation logic and HIP
applies Hoare rules to programs and uses SLEEK to discharge the associated proof obligations. We
proceeded as follows:
\begin{enumerate}[(1)]
\item We developed an equational solver over the sophisticated fractional share model of
    Dockins \emph{et al.} \cite{dockins09:sa}.  Permissions can be existentially or universally
    quantified and arbitrarily related to permission constants.
\item We integrated our equational solver over shares into SLEEK to handle fractional
    permissions on separation logic assertions (\emph{e.g.}, points-to, etc.). We believe that
    SLEEK is the first automatic entailment checker for separation logic that can handle a
    sophisticated share model (although some other tools can handle simpler share models).
\item We developed an encoding of barrier definitions (diagrams) in SLEEK, which now
    automatically verifies the side conditions from \S\ref{sec:bardefs}.
\item We modified HIP to recognize barrier definitions (whose side conditions are then verified
    in SLEEK) and barrier calls using the Hoare rule from Figure~\ref{fig:hoare_seq}.
\end{enumerate}
Next we describe our equational solver for the Dockins \emph{et al.} share model before giving a
more technical background to the HIP/SLEEK system and describing our modifications to it in detail.
Most of the technical work occurred in developing the equational solver and its integration with
the rest of the separation logic entailment procedures in SLEEK.  Once SLEEK understood fractional
permissions, checking the validity side conditions on barriers was quite simple.

\subsection{Decision Procedure for Shares}

SLEEK discharges the heap-related proof obligations but relies on external decision procedures for
the pure logical fragments it extracts from separation logic formulae. For example, SLEEK utilizes
Omega for Presburger arithmetic, Redlog for arithmetic in $\mathbb{R}$, and MONA for monadic
second-order logic.  Adding fractional permissions required an appropriate
equational decision procedure for fractional shares.

Decision procedures for simple fraction share models such as rationals between $0$
and $1$ need only solve systems of linear equations.  The more sophisticated fractional share
model of Dockins \emph{et al.} \cite{dockins09:sa} requires a more sophisticated solver.

Dockins \emph{et al.} represent shares as binary trees with boolean-valued leaves.  The full share
$\fullp$ is a tree with one true leaf $\bullet$ and the empty share $\emptyp$ is a tree with one
false leaf $\circ$.  The left-half share $\lhalfp$ is a tree with two leaves, one true and one
false: $\ltree$; similarly, the right-half share $\rhalfp$ is a tree with two leaves, one false and
one true: $\rtree$.  The trees can continue to be split indefinitely: for example, the right half
of $\lhalfp$ is $\Tree [ [ $\circ$ $\bullet$ ] $\circ$ ] $.  Joining is defined by structural
induction on the shape of the trees with base cases $\circ \oplus \circ = \circ$, $\bullet
\oplus \circ = \bullet$, and $\circ \oplus \bullet = \bullet$ (emphasis: $\oplus$ is partial). When
two trees do not have the same shape, they are unfolded according to the rules $\bullet \cong \Tree
[ $\bullet$ $\bullet$ ] $ and $\circ \cong \Tree [ $\circ$ $\circ$ ] $; for example:
\[
\Tree [ [ $\circ$ $\bullet$ ] $\circ$ ] ~ \oplus ~ \Tree [ [ $\bullet$ $\circ$ ] [ $\circ$ $\bullet$ ] ] \quad = \quad
\Tree [ [ $\circ$ $\bullet$ ] [ $\circ$ $\circ$ ] ] ~ \oplus ~ \Tree [ [ $\bullet$ $\circ$ ] [ $\circ$ $\bullet$ ] ] \quad = \quad
\Tree [ [ $\bullet$ $\bullet$ ] [ $\circ$ $\bullet$ ] ] \quad = \quad
\Tree [ $\bullet$ [ $\circ$ $\bullet$ ] ]
\]

SLEEK takes a formula in separation logic with fractional shares and extracts a specialized formula
over \textbf{strictly positive} shares whose syntax is as follows:
\[
\phi ~::=~ \exists v.\phi ~|~
           \phi_1 \vee \phi_2 ~|~
           \phi_1\wedge \phi_2 ~|~
           v_1 \oplus v_2 = v_3 ~|~
           v_1 = v_2 ~|~
           v = \chi
\]
Our share formulae $\phi$ contain share variables $v$, existentials $\exists$, conjunctions
$\wedge$, disjunctions $\vee$, join facts $\oplus$, equalities between variables, and assignments
of variables to constants $\chi$.  The tool also recognizes $v
{\in}[\chi_1,\chi_2]$, pronounced ``$v$ is bounded by $\chi_1$ and $\chi_2$'', which is
semantically equal to:
\[
((v = \chi_1) \vee (\exists v'.~ \chi_1 \oplus v' = v)) ~ \wedge ~
((v = \chi_2) \vee (\exists v''.~ v \oplus v'' = \chi_2))
\]
Disjunctions are needed because share variables can only be instantiated with positive shares:
$\forall v. \not \hspace{-1.5pt} \exists v'. v \oplus v' = v$.  Handling
bounds checks ``natively'' rather than compiling them into semantic definitions increases efficiency by reducing the number of existentials
and disjunctions.

SLEEK asks the solver questions of the following forms:
\begin{enumerate}[(1)]
\item (UNSAT) Is a given formula $\phi$ unsatisfiable?
\item ($\exists$-ELIM) Given a formula of the form $\exists v.~\phi(v)$, is there a unique
    constant $\chi$ such that $\exists v.~ \phi(v)$ is equivalent to $\phi(\chi)$?
\item (IMPL) Given two formulae $\phi_1$ and $\phi_2$, does $\phi_1$ entail $\phi_2$?
\end{enumerate}
Our solver is sound but incomplete.  However, it is \emph{complete enough} to help SLEEK check a
wide variety of entailments involving fractional permissions, including all of those in the example
from Figure \ref{fig:ex_code}. \hide{ We have settled for a procedure that soundly approximates the
solution, leaving open the possibility of further improvements with the observation that in our
experiments this mechanism proved to be more than enough.}

\hide{(highly)}

All of these questions can be reduced to solving a series of constraint systems
whose equations are of the form $v_1 \oplus v_2 = v_3$, $v {\in}[\chi_1,\chi_2]$, and $v = \chi$.
Solving constraint systems in separation algebras (\emph{i.e.}, cancellative partial commutative
monoids) is not as straightforward as it might seem because many of the traditional algebraic
techniques do not apply.  Our lightweight constraint solver finds an overapproximation to the
solution, returning either (a) the constant \texttt{UNSAT} or (b) for each variable $v_i$ either an
\textbf{assignment} $v_i = \chi$ or a \textbf{bound} $v_i \in [\chi_1, \chi_2]$ such that:
\begin{iteMize}{$\bullet$}
\item (FALSE) If the algorithm returns \texttt{UNSAT}, then the formula is unsatisfiable.  The
    algorithm will return \texttt{UNSAT} if it discovers a bound whose ``lower value'' is
    higher than its ``upper value'', or if it discovers a falsehood (\emph{e.g.}, after
    constant propagation one of the equations becomes $\fullp \oplus \fullp = \fullp)$.
\item (COMPLETE) All solutions to the system (if any) lie within the bounds.
\item (SAT-PRECISE) A solution is \emph{precise} when all variables are given
    assignments.  If a solution is precise, then the formula is satisfiable.
\end{iteMize}
SLEEK queries are given in share formulae that must be transformed into the equational systems
understood by our constraint solver.  To do this transformation, first we put the relevant formulae
into disjunctive normal form (DNF).  Each disjunct becomes an independent system of equations.
Given one disjunct we form this system by simply treating each basic constraint (\emph{i.e.},
$v=v'$, $v=\chi$, $v \in [\chi_1,\chi_2]$, and $v_1 \oplus v_2 = v_3$) as an equation.  Our solver
approximates each system independently and can then answer SLEEK's questions as follows:
\begin{iteMize}{$\bullet$}
\item (UNSAT): Return \texttt{False} when
the algorithm returns \texttt{UNSAT} for each constraint system obtained from the formula;
otherwise return \texttt{True}.
\item ($\exists$-ELIM): If the variable $v$ has the same assignment in all constraint systems
    derived from the DNF, then return that value.  It is sound to substitute that value for $v$
    and eliminate the existential.  (If the formula is satisfiable, then that is the unique
    assignment that makes it so; if the formula is false then after the assignment it will
    still be false.) \hide{ In either case, we can substituted that value to eliminate the
    existential without affecting the truth of the formula.  the restriction that the same
    value be a solution for all systems is stronger than generally required however we impose
    it due to the specifics of the interaction with SLEEK.}
\item (IMPL): Return \texttt{True} only when either:
\begin{iteMize}{$-$}
  \item the solver returns \texttt{UNSAT} for all systems derived from the antecedent
  \item the solver returns a precise solution for each system of equations derived from the
      antecedent, and the solver also returns \textbf{the same} precise solution for at
      least one of the consequent systems.
\end{iteMize}
\end{iteMize}

The constraint solver works by eliminating one class of constraints at a time:
\begin{enumerate}[(1)]
  \item First we substitute $v = \chi$ constraints into the remaining equations.
  \item We handle $\oplus$ constraints with exactly one variable as follows:
   \begin{iteMize}{$\bullet$}
     \item $\chi_1 \oplus \chi_2 = v$: we check if the join is defined, and if so
         substitute the sum for $v$ in the remaining equations; otherwise, we return
         \texttt{UNSAT}.
    \item $\chi_l \oplus v = \chi_r$ or $v \oplus \chi_l = \chi_r$: we check if $\chi_r$
        contains $\chi_l$, and if so substitute the difference $\chi_r-\chi_l$ for $v$ in
        the remaining equations; otherwise return \texttt{UNSAT}.  (``$-$'' has the
        property that if $\chi_1 - \chi_2 = \chi_3$ then $\chi_3 \oplus \chi_2 = \chi_1$).
	\end{iteMize}  	
  \item Constraints involving constants ($\chi_1 \oplus \chi_2 = \chi_3$ and $ \chi
      {\in}[\chi_1,\chi_2]$) are dismissed if the equality/inequalities hold; otherwise return
      \texttt{UNSAT}.
  \item We attempt to dismiss certain kinds of unsatisfiable systems via a consistency check as
      follows.  We first compute the transitive closure of variable substitutions, resulting in
      facts of the form $v_1 \oplus \ldots \oplus v_n \oplus \chi_1 \oplus \ldots \oplus \chi_m
      = \chi$.  Nonempty shares \textbf{cannot} join with themselves.  Therefore, if the $v_i$
      contain duplicates we return \texttt{UNSAT}.  We also return \texttt{UNSAT} if the
      constants $\chi_i$ do not join or if $\chi$ does not contain $\chi_1 \oplus \ldots \oplus
      \chi_m$.
  \item Variables in the remaining constraints are given initial domains of
  ($\emptyp,\fullp$).
  \item Each $\in$ constraint is used to restrict the domain of its corresponding variable.
  \item At this point only $a_1 \oplus a_2 = a_3$ constraints involving at least two variables
      remain.  The algorithm then proceeds by iteratively selecting an equation, checking it
      for consistency, and then refining the associated domains via a forward and backward
      propagation.  The algorithm iterates until either a fixpoint is reached or a consistency
      check fails.  To check an equation for consistency, the algorithm verifies that:
      \begin{iteMize}{$\bullet$}
        \item for each variable, the lower bound is less than the upper bound
   		\item the current lower bounds of the LHS variables join together
   		\item the join of the LHS lower bounds is below the RHS upper bound
   		\item the join of the LHS upper bounds is above the RHS lower bound
     \end{iteMize}
      Forward propagation consists of (Fa) lowering the upper bound of the RHS by intersecting
      away any subtree that does not appear in the upper bounds of the LHS, and (Fb) increasing
      the lower bound of the RHS by unioning all subtrees present in the lower bounds in the
      LHS.  Backwards propagation consists of (Ba) lowering the upper bounds of the LHS by
      intersecting away any subtree that does not appear in the upper bound on the RHS.
      Increasing the lower bounds of the LHS (Bb) is trickier since we do not know \emph{which
      operand} should be increased.  There are several possibilities we could have taken, but
      we selected the simplest: we simply leave the bounds as they were unless one of the
      operands has been determined to be a constant, in which case we can calculate exactly
      what the lower bound for the other variable should be.  This solution is can lead to
      overapproximation, but a more refined solution would require a performance cost, which
      did not seem warranted by our experiments.  After each forward/backwards propagation, if
      we have refined a domain to a single point, the variable is substituted for a constant
      value of that point in the remaining equations.

      \qquad Once we reach a fixpoint, the resulting variable bounds represent an over
      approximation of the solution.
\end{enumerate}

\hide{  Note that during the backward propagation step, the lower bounds of the operands are
refined by
  adding subtrees of the result lower bound however there is only a constraint of the joining of the
  added subtrees, not on the individual subtrees. This leads to the solving procedure needing either
  to pick a splitting or to explore all viable splittings. We have chosen to forego this
  particular refinement for situations where there is more than one splitting(i.e. both operands
  being variables). Due to this restriction we are able to provide only an approximation of all the
  possible solutions. A more refined solution representation could solve this issue albeit
  at a performance cost which did not seem warranted by our experiments.}

\subsection{An introduction to SLEEK}

SLEEK checks entailments in separation logic~\cite{nguyenVMCAI}.  The antecedent may cover more of
the heap than the consequent, in which case SLEEK returns this residual heap together with the pure
portion of the antecedent.  SLEEK can also discover instantiations for certain existentials in the
consequent, a feature that we elide here; details may be found in Chin \emph{et al.} \cite{Chin2010}.

\hide{(\emph{e.g.} for a linked list predicate, $\mathsf{self}$ is the pointer to the head of the
list)}

One of SLEEK's strong points is that it allows user-defined inductive predicates. Predicates are
defined as separation formulae that describe the shape of data structures and associated properties
(\emph{e.g.}, list length, tree height, and bag of values contained in a list).  SLEEK uses the
keyword $\mathsf{self}$ as a pointer variable to the current object.  Predicate invariants can
increase the precision of the verification (\emph{e.g.}, length $\geq$ 0).  An invariant for a
predicate instance has two parts: a pure formula describing arithmetic constraints
on the arguments and the set of non null pointer arguments (\emph{e.g.} the outward pointer for a
list segment).

Figure \ref{SLEEKlang} gives an outline of the specification language accepted by SLEEK with our
extensions for the fractional permissions.  The system accepts disjunctive separation logic
formulae ($\Phi$) with both heap ($\kappa$) and pure ($\pi$) constraints; we denote the disjunction
by $\bigvee$. The syntax allows richer structures as well, \emph{e.g.} directed case
analysis and staged formulae (corresponding to the $\Phi ~[Q]$ form) as described
in \cite{FM11:Gherghina}. \hide{, and
uniformly handling pre/post conditions ; we use the latter feature to enable a program verifier like HIP
to verify barrier calls.}
Staged formulae help split implication proofs
into stages such that redundant proving is eliminated and ensure that key constraints are proven early, \emph{e.g.}, before applying 
case analysis.
In order to prove that $\Phi ~[Q]$ holds, $\Phi$ is proven before $Q$ is proven.

At the core of a separation logic formula are the heap constraints.  Heap constraints are heap
node descriptions connected by the separating conjunction.  A node is either an instance of a data structure
or an instance of a user-defined predicate.  Here we use the same notation for both cases:
$\hdata{v}{c}{v^*}$, where $\term{v}$ is the pointer to the structure, $\term{c}$ is the
data structure type or predicate name, and $\term{v}^*$ is the list of arguments (either
predicate arguments for predicate instances or field values for data structures).  Separation
logic formulae can also contain pure constraints over several domains: arithmetic, bag/list,
etc.  For brevity we discuss only arithmetic constraints in this presentation.


\begin{figure}
\begin{center}
\begin{minipage}{33pc}
\begin{frameit}
\vspace{-4mm}
\footnotesize
\[
\begin{array}{lrcl}
\mathrm{Predicate} & \mathsf{spred} &::=& [\mathsf{self}{::}]\view{\term{c}}{\term{v}^*}
\equiv \mathsf{Q}~[{\bf inv}~(\pure,\term{v}^*)] \\
\mathrm{Formula} & \mathsf{Q} &::=& \mathsf{R} \mid
\mathsf{R}\vee\mathsf{Q}\\ & \mathsf{R} &::=&  \code{case} \{ [\pure {\rightarrow} \mathsf{Q}]^+
 \} \mid \hide{[\term{w}^*]} ~\constr~ [\mathsf{Q}] \\
& \constr &::=& \bigvee_i (\exists \term{v_i}^* \cdot (\heap_i \wedge \pure_i \wedge \tau_i))
\qquad\qquad \Delta ~::=~ {\bigvee}_i (\heap_i \wedge \pure_i \wedge \tau_i)\\
\mathrm{Frac~form.}&\tau & ::=& v_{f} \oplus v_{f} = v_{f} ~|~ v
{\in}[\chi,\chi] ~|~ v = \chi ~|~ \tau \wedge \tau \\
\mathrm{Pure~form.} & \pure &::=& \ptr \wedge \pconstr \\
\mathrm{Pointer~form.} & \ptr &::=& \term{v} = \term{v} \mid v=\nil \mid
\term{v}\neq \term{v} \mid \term{v} \neq \nil \mid \ptr \wedge \ptr  \\
\mathrm {Heap~form.} &\heap &::=& \emp ~|~ \hfdata{v}{c}{v_{f}}{v^*} \hide{~|~
 \hfpred{v}{p}{v_{f}}{v^*}} ~|~ \heap \sep \heap \hide{~|~ \heap_1 \mw \heap_2}
\\
 \mathrm{Presburger~arith.} & \phi &::=& \mathsf{arith} \mid \phi \wedge
 \phi \mid \phi \vee \phi \mid \neg \phi \mid \exists \term{v} \cdot \phi \mid \forall \term{v} \cdot \phi \\
& \mathsf{arith} &::=& \mathsf{a} = \mathsf{a} \mid \mathsf{a} \neq \mathsf{a} \mid
			\mathsf{a} < \mathsf{a} \mid \mathsf{a} \leq \mathsf{a}  \\
& \mathsf{a} &::=& \term{z} \mid \term{v} \mid \term{z} \times \mathsf{a} \mid \mathsf{a} +
\mathsf{a} \mid -\mathsf{a} \mid \myit{max} (\mathsf{a}, \mathsf{a}) \mid
\myit{min}(\mathsf{a}, \mathsf{a}) \\
\end{array}\] \[
\begin{array}{lcl}
\myit{where} \hide{&p&\myit{is a predicate name};\\}
&v,w&\myit{are variable names};\\
& c& \myit{is a data type name or a predicate name};\\
& \term{z}& \myit{is an integer constant};\\
& \tau & \myit{represents the fractional permission constraints}\\
& \chi & \myit{represents constant fractional shares}\\
 \\[-6pt]
\end{array}
\]
\end{frameit}
\vspace{-3mm}
\caption{The Specification Language with Fractional Permissions.}
\label{SLEEKlang}
\end{minipage}
\end{center}
\end{figure}

The syntax in Figure \ref{SLEEKlang} contains two new extensions to SLEEK's language. First, heap
node descriptions can contain permission annotations for fractional ownership.  A
heap node partially-owned with share $v_{f}$ is indicated by $\hfdata{v}{c}{v_{f}}{v^*}$.  If $c$
denotes a predicate, then the notation $\hfpred{v}{c}{v_{f}}{v^*}$ indicates that $v$ points to a memory region whose
shape is described by the definition of $c$. Furthermore this notation denotes that all heap nodes
abstracted by this predicate instance are owned with permission $v_{f}$
(\emph{e.g.}, in a $\lhalfp$-owned list, each list cell is owned $\lhalfp$).  A
node/predicate without a permission annotation indicates full ownership.  The second extension
enables the expression of constraints over fractional permission variables using the syntax
$v_{f1} \oplus v_{f2}=v_{f3}$, $v {\in} [\chi_1, ~\chi_2]$, $v_1=v_2$, and $v=\chi$.

\subsection {SLEEK entailment background}

The core of the SLEEK entailment works by algorithmically discharging the heap obligations and then
referring any remaining pure constraints to other provers.  SLEEK discharges heap obligations in
three ways: heap node matching, predicate folding, and predicate unfolding.  To guarantee
termination, SLEEK ensures that each predicate fold or unfold must be immediately followed by a match,
and that no two fold operations for the same predicate are performed in order to match one node.
These restrictions ensure that each successful fold, unfold, and match operation decreases the number of RHS nodes.

Entailments in SLEEK are written as follows:
$\entailK{\heap}{V}{\D_A}{\mathsf{Q}_C}{\D_R}$, which is shorthand for
$\sm{\entailH{\heap \sep \D_A}{\exists V {\cdot} (\heap \sep \mathsf{Q}_C)}{\D_R}}$.  The entailment
checks whether the consequent heap nodes $\small{\mathsf{Q}_C}$ are covered by heap nodes in
antecedent $\sm{\D_A}$, and if so, SLEEK returns the residual heap $\sm{\D_R}$, which consists of the
antecedent nodes that were not used to cover $\sm{\mathsf{Q}_C}$. The implementation
performs a proof search and thus returns a set of residues. For simplicity, assume that only one
residue is computed. In the entailment, $\sm{\heap}$ is the history of nodes from the antecedent
that have been used to match nodes from the consequent, $\sm{V}$ is the list of existentially
quantified variables from the consequent.  Note that $\kappa$ and $\sm{V}$ are discovered
iteratively: entailment checking begins with $\sm{\heap=\emp}$ and $\sm{V=\emptyset}$.

The initial system behavior was described in detail in \cite{nguyenVMCAI,Chin2010,FM11:Gherghina}.
The main rules for matching, folding, unfolding, and discharging of pure
constraints are given here. The initial main entailment checking rules are given in Fig~\ref{fig.nonDentail}.
Later we show how we modified these rules to accommodate fractional shares.

\begin{figure}
\begin{center}
\begin{minipage}{25pc}
\begin{frameit}
\[
\begin{array}{c}
\xpure(\emp) ~\equiv~ (\true,\emptyset)\\
[10pt]\afrac{\myit{IsData}(c)}
{\xpure(\hdata{p}{c}{v^*}) ~\equiv~(p{\neq}0;\{p\})}\\
[20pt]\afrac{
\myit{IsPred}(c) \qquad (\view{c}{v^*} \equiv \mathsf{Q}~\code{inv}~(\pi_1,\pi_2)) \in \myit{P}\\
} {\xpure(\hpred{p}{c}{v_p^*}) \equiv ([p/\mathsf{self},v_p^*/v^*]\pi_1,[p/\mathsf{self},v_p^*/v^*]\pi_2)}\\
[20pt]\afrac{\xpure(\heap_1)\equiv (f_1,s_1) \qquad \xpure(\heap_2)\equiv (f_2,s_2)
}{\xpure (\heap_1\sep\heap_2) \equiv (f_1\wedge f_2 , s_1 \cup s_2) }
\end{array}
\]
\end{frameit} \caption{\myit{X\!Pure}~:~Translating to Pure
Form}\label{fig.map}
\end{minipage}
\end{center}
\end{figure}

Entailment between separation formulae is reduced to entailment between pure formulae by matching
heap nodes in the RHS to heap nodes in the LHS (possibly after a fold/unfold).  Once the RHS is
pure, the remaining LHS heap formula is soundly approximated to a pair of pure formula and set of
disjoint pointers by function {\small{$\xpure$}} as defined in Fig~\ref{fig.map}. The functions
{\small{\myit{IsData}(c)}} and {\small{\myit{IsPred}(c)}} decide respectively if $c$ is a data
structure or a predicate.  The procedure successively pairs up heap nodes that it proves are
aliased. SLEEK keeps the successfully matched nodes from the antecedent in $\sm{\heap}$ for better
precision in the next iteration.

\begin{figure*}[thb]
\begin{center}
\begin{minipage}{14.6cm}
\begin{frameit}\vspace*{-5mm}
\[
\!
\infer[]{
\entailVV{\heap_1{\wedge}\pure_1}{\pure_2}{(\heap_1{\wedge}\pure_1)}
}{
\begin{array}{c}
\mathrm{EMP}\\
(\rho, S){=}\xpure(\heap_1{\sep}\heap)\\
\rho{\wedge}(\forall x,y \in S \cdot x\neq y)
\!\!{\implies}\!\!{{\exists} \myit{V}{\cdot}\pure_2}\!\!
\end{array}
}
\quad
\infer[]
{\entailVV{\hpred{p_1}{c}{v^*_1}{\sep}\heap_1{\wedge}\pure_1}
{(\hpred{p_2}{c}{v^*_2}{\sep}\heap_2{\wedge}\pure_2)}{\D} }
{\begin{array}{c}
 \mathrm{MATCH}\\
\mathsf{fst}(\xpure(\hpred{p_1}{c}{v^*_1}{\sep}\heap_1{\sep}\heap)){\wedge}\pure_1\!\!{\implies}\!\!p_1{=}p_2 \\
\entailK{\heap{\sep}\hpred{p_1}{c}{v^*_1}}{V}{\heap_1{\wedge}\pure_1}{\heap_2{\wedge}\pure_2{\wedge}(\bigwedge_i(v^i_1=v^i_2))}{\D}
\end{array}}
\]

\[
\infer[]
{\entailVV{\hdata{p_1}{c_1}{v^*_1}{\sep}\heap_1{\wedge}\pure_1}
{(\hpred{p_2}{c_2}{v^*_2}{\sep}\heap_2{\wedge}\pure_2)}{\D}}
{\begin{array}{c}
 \mathrm{FOLD}\\
\myit{IsPred}(c_2){\wedge}\myit{IsData}(c_1) \quad \hpredhead{c_2}{v^*}{\equiv}\mathsf{Q}
\in \myit{P}\\
\!\!
\mathsf{fst}(\xpure(\hdata{p_1}{c_1}{v^*_1}{\sep}\heap_1{\sep}\heap)) \wedge \pure_1
\!\!{\implies}\!\!p_1{=}p_2\\
\entailK{\heap}{\emptyset}{\hdata{p_1}{c_1}{v^*_1}{\sep}\heap_1{\wedge}\pure_1}
{[p_1/\mathsf{self},v_1^*/v^*]\mathsf{Q}}{\Delta^r}\\
\entailK{\heap}{V}{\Delta^r}{(\heap_2{\wedge}\pure_2)}{\D}
\end{array}}
\]

\[
\infer[]
{\entailVV{\hpred{p_1}{c_1}{v^*_1}{\sep}\heap_1{\wedge}\pure_1}
{(\hdata{p_2}{c_2}{v^*_2}{\sep}\heap_2{\wedge}\pure_2)}{\D}}
{\!\begin{array}{c}
\mathrm{UNFOLD}\\
\myit{IsPred}(c_1){\wedge}\myit{IsData}(c_2) \quad \hpredhead{c_1}{v^*}{\equiv}\mathsf{Q} \in
\myit{P} \\
\mathsf{fst}(\xpure(\hpred{p_1}{c_1}{v^*_1}{\sep}\heap_1{\sep}\heap))\wedge\pure_1\!\!{\implies}\!\!p_1{=}p_2\\
\Delta_Q = \mathsf{to\_disjunct}(\mathsf{Q})\\
\entailVV{[p_1/\mathsf{self},v_1^*/v^*]\Delta_Q{\sep}\heap_1{\wedge}\pure_1}
{(\hdata{p_2}{c_2}{v^*_2}{\sep}\heap_2{\wedge}\pure_2)}{\D}
\end{array}}
\]
\end{frameit}\vspace*{-3mm}
\caption{Separation Constraint Entailment}\label{fig.nonDentail}
\vspace*{-3mm}
\end{minipage}
\end{center}
\end{figure*}

All three heap reducing steps start by establishing that there is a heap node on the LHS of the
entailment that is aliased with the RHS heap node that is to be reduced ($p_1=p_2$). In order to
prove the aliasing, the LHS heap together with the previously consumed nodes are approximated to a
pure formula, and together with the LHS pure formula the $p_1=p_2$ implication is checked.
Similarly, when a match occurs (rule $\mathrm {MATCH}$), equality between node arguments needs to be 
proven.
\hide{
 all the free arguments of the matched node from the RHS are
instantiated and those instantiations are moved to the LHS.  The free RHS arguments are bound to
the corresponding argument in the LHS and the equalities are moved to the LHS.

These instantiations are particularly useful for the entailments constructed by program verifiers
like HIP when verifying a method call, because free variables in the RHS are usually variables from
the precondition of the callee and these bindings are useful in subsequent program verification
steps. By moving the bindings to the LHS, they can be carried forward in the residual state and thus
returned to the program verifier. This process is formalized by the function \sm{\myit{freeEqn}}
below, where \sm{V} is the set of existentially quantified variables:
\[
\begin{array}{l}
\myit{freeEqn}([u_i/v_i]_{i=1}^{n}, V) \equiv \bigwedge_{i=1}^{n}(\btt{if}~v_i \in V ~\btt{then}~ \true
~\btt{else}~v_i{=}u_i)
\end{array}
\]}

Unfold and fold operations handle inductive predicates in a deductive manner.  SLEEK can unfold a
predicate instance that appears in the LHS if the unfolding exposes a heap node that matches
immediately with a node in the RHS. Similarly, several LHS nodes can be folded into a predicate
instance if the resulting predicate instance can be immediately matched with a RHS node.
Well-formedness conditions imposed on the predicate definitions ensure that after a fold or unfold
a matching always takes place; these conditions have been elided for this presentation.  The unfold
rule presents the replacement of a predicate instance in which the predicate definition is reduced to a
disjunctive form and in which the arguments have been substituted.  The fold step requires the LHS
to entail the predicate definition. The residue of this entailment is then used as the new LHS for
the rest of the original entailment. \hide{The folding rule needs to carefully deal with the
instantiations: some move to the LHS while others still need to be proven and thus are moved to the
RHS.  This partitioning of the pure constraints is done by the split function.}  For a more
detailed explanation of the SLEEK entailment process, see Chin \emph{et al.} \cite{Chin2010}.

\subsection{Entailment Procedure for Separation Logic with Shares}

\begin{figure}
\begin{center}
\begin{minipage}{32pc}
\begin{frameit}
\vspace{-5mm}
\[
\begin{array}{c}
\fxpure(\emp,\tau) ~\equiv~ (\true,\emptyset)\\
[5pt]
\afrac{\myit{IsData}(c) \qquad \tau \Rightarrow v_f=cs}
{\fxpure(\hfdata{p}{c}{v_f}{v^*},\tau) ~\equiv~(p{\neq}0;\{(p,cs)\})}\\
[15pt]
\afrac{\fxpure(\heap_1,\tau)\equiv (f_1,s_1) \qquad \fxpure(\heap_2,\tau)\equiv (f_2,s_2)
}{\fxpure (\heap_1\sep\heap_2,\tau) \equiv (f_1\wedge f_2 , s_1 \cup s_2) }\\
[15pt]
\afrac{
\myit{IsPred}(c) \qquad (\view{c}{v^*} \equiv \mathsf{Q}~\code{inv}~(\pi_1,\pi_2)) \in \myit{P}\\
\tau \Rightarrow v_f=cs \qquad \pi_1'=[p/\mathsf{self}]\pi_1 \qquad \pi_2'=\{\forall v \in \pi_2,
([p/\mathsf{self}]v,cs)\}\\
} {\xpure(\hfpred{p}{c}{v_f}{v^*},\tau) \equiv (\pi_1',\pi_2')}
\end{array}
\]
\end{frameit} \caption{\myit{$\fxpure$: $\xpure$ with shares}}\label{fig.map_wshares}
\end{minipage}
\end{center}
\end{figure}

\begin{figure*}[thb]
\begin{center}
\begin{minipage}{12.6cm}
\begin{frameit}\vspace*{-5mm}

\[
\infer[]
{\entailVV{\hfdata{p_1}{c_1}{f_1}{v^*_1}{\sep}\heap_1{\wedge}\pure_1\wedge \tau_1}
{(\hfpred{p_2}{c_2}{f_2}{v^*_2}{\sep}\heap_2{\wedge}\pure_2\wedge\tau_2)}{\D}}
{\begin{array}{c}
 \mathrm{FOLD}\\
\myit{IsPred}(c_2){\wedge}\myit{IsData}(c_1) \quad
\hpredhead{c_2}{v^*}{\equiv}\mathsf{Q} \in \myit{P} \\
\!\!
\mathsf{fst} (\fxpure(\hfdata{p_1}{c_1}{f_1}{v^*_1}{\sep}\heap_1{\sep}\heap, \tau_1)) \wedge \pure_1
\!\!{\implies}\!\!p_1{=}p_2\\
\mathsf{Q}'=\mathsf{set\_shares}([p_1/\mathsf{self},v^*_1/v]\mathsf{Q}, f_2)\\
\entailK{\heap}{\emptyset}{\hfdata{p_1}{c_1}{f_1}{v^*_1}{\sep}\heap_1{\wedge}\pure_1{\wedge}\tau_1}
{\mathsf{Q}'}{\Delta^r}\\
\entailK{\heap^r}{V}{\D^r}{(\heap_2{\wedge}\pure_2{\wedge}\tau_2)}{\D}
\end{array}}
\]

\[
\infer[]
{\entailVV{\hfpred{p_1}{c_1}{f_1}{v^*_1}{\sep}\heap_1{\wedge}\pure_1{\wedge}\tau_1}
{(\hfdata{p_2}{c_2}{f_2}{v^*_2}{\sep}\heap_2{\wedge}\pure_2{\wedge}\tau_2)}{\D}}
{\!\begin{array}{c}
\mathrm{UNFOLD}\\
\hpredhead{c_1}{v^*}{\equiv}\mathsf{Q} \in \myit{P} \quad
\myit{IsPred}(c_1){\wedge}\myit{IsData}(c_2) \\
\mathsf{fst}
(\fxpure(\hfpred{p_1}{c_1}{f_1}{v^*_1}{\sep}\heap_1{\sep}\heap,\tau_1))\wedge\pure_1\!\!{\implies}\!\!p_1{=}p_2\\
\mathsf{Q}' = \mathsf{set\_shares}([p_1/\mathsf{self},v_1^*/v^*]\mathsf{Q},f_1) \\
\Delta_Q = \mathsf{to\_disjunct}(\mathsf{Q}')\\
\entailVV{\Delta_Q{\sep}\heap_1{\wedge}\pure_1{\wedge}\tau_1}
{(\hfdata{p_2}{c_2}{f_2}{v^*_2}{\sep}\heap_2{\wedge}\pure_2{\wedge}\tau_2)}{\D}
\end{array}}
\]
\end{frameit}\vspace*{-3mm}
\caption{Folding/Unfolding in the presence of shares}\label{fig.shares_fold} \vspace*{-3mm}
\end{minipage}
\end{center}
\end{figure*}

Adding fractional permissions required several modifications to the entailment process.
\begin{iteMize}{$\bullet$}
 \item {\bf Empty heap}. In a separation logic without shares, whenever $(\exists a,b.
     \heapn{x}{}{a} \sep \heapn{y}{}{b})$ then $x\neq y$.  In SLEEK, this fact is captured in
     the $\mathsf{EMP}$ rule, which tries to prove the pure part of the consequent after
     enriching the antecedent pure formula with pure information collected from the previously
     consumed heap and the remaining LHS heap.  It extracts both the invariants of the heap
     nodes and constructs a formula that ensures that all pointers in the heap are
     distinct.

     Introducing fractional permissions requires the relaxation of this
     constraint because $\exists a,b. \heapn{x}{x_f}{a} \sep \heapn{y}{y_f}{b}$
     implies $x\neq y$ only if the $x_f$ and $y_f$ shares overlap.  We changed the $\xpure$ function
     to return a pair of a pure formula, and pairs of pointers and associated fractional shares. The new version
     of $\xpure$ allowed the $\mathsf{EMP}$ rule to be rewritten to enforce inequality
     only between pointers that have conflicting shares:
     {\[
	 \infer[\mathsf{EMP}]{
	 \entailVV{\heap_1 \wedge \pure_1 \wedge \tau_1}{\pure_2}{(\heap_1 \wedge \pure_1 \wedge \tau_1)}
	 }{
	 \begin{array}{c}
	 (\rho, S){=}\fxpure(\heap_1{\sep}\heap, \tau)\\
	 \rho{\wedge}(\forall (x,x_f),(y,y_f) \in S, (\neg \exists z \cdot x_f \oplus y_f = z) \cdot x\neq
	 y) \!\!{\implies}\!\!{\exists \myit{V}{\cdot}\pure_2}\!\!
	 \end{array}
	 }
     \]}

  \item {\bf Folding/unfolding}. By convention, all the heap nodes
  		abstracted by a predicate instance are owned with the same fractional permission as the
      predicate instance. Therefore, unfolding a node first replaces the permissions of the nodes
      in the predicate definition with the permission of that LHS node. Then the updated predicate
      definition replaces the predicate instance. Similarly, folding a node replaces the permissions of all
      nodes in the definition with the permission of that RHS node before trying to entail the
      predicate definition. The $\mathsf{set\_shares}(\mathsf{Q},v)$ function sets the permissions of all heap
      nodes in $\mathsf{Q}$ to $v$. The new set of rules is shown in Figure \ref{fig.shares_fold}.

  \item {\bf Matching}. In order to properly handle a match in the presence of fractional
      shares, the entailment process needs to (a) reduce both LHS and RHS nodes entirely, or
      (b) split the LHS node and reduce one side, or (c) split the RHS and reduce one side.
  \[
	  \infer[\mathsf{FULL}\textrm{-}\mathsf{MATCH}\textrm{ (a)}]
	{\entailVV{\hfpred{p_1}{c}{f_1}{v^*_1}{\sep}\heap_1 \wedge \pure_1 \wedge \tau_1}
	{(\hfpred{p_2}{c}{f_2}{v^*_2}{\sep}\heap_2 \wedge \pure_2 \wedge \tau_2)}{\D} }
	{\begin{array}{c}
	\big( \mathsf{fst}(\fxpure(\hfpred{p_1}{c}{f_1}{v^*_1}{\sep}\heap_1,
	\tau_1)) \wedge \pure_1 \big) ~~ \implies ~~ p_1=p_2 \\
    \heap' = \heap{\sep}\hfpred{p_1}{c}{f_1}{v^*_1}\\
    \rho = f_1{=}f_2\wedge(\bigwedge_i(v^i_1=v^i_2))\\
	\entailK{\heap'}{V}{\heap_1 \wedge \pure_1 \wedge \tau_1}
	   {\heap_2 \wedge \pure_2 \wedge \tau_2\wedge \rho}{\D}
	\end{array}}
  \]
\\
  \[
	  \infer[\begin{array}{l} \mathsf{LEFT}\textrm{-} \\ \mathsf{SPLIT}\textrm{-} \\ \mathsf{MATCH}\textrm{ (b)} \end{array}]
	{\entailVV{\hfpred{p_1}{c}{f_1}{v^*_1}{\sep}\heap_1{\wedge}\pure_1{\wedge}\tau_1}
	{(\hfpred{p_2}{c}{f_2}{v^*_2}{\sep}\heap_2{\wedge}\pure_2{\wedge}\tau_2)}{\D} }
	{\begin{array}{c}
	\mathsf{fst}(\fxpure(\hfpred{p_1}{c}{f_1}{v^*_1}{\sep}\heap_1,
	\tau_1)){\wedge}\pure_1 ~~ \implies ~~ p_1{=}p_2 \\
	\tau_1'=\tau_1 \wedge f_{c1}\oplus f_{r1} = f_1 \\
	\heap' = \heap{\sep}\hfpred{p_1}{c}{f_{c1}}{v^*_1}\\
	\rho = f_{c1}{=}f_2{\wedge}(\bigwedge_i(v^i_1=v^i_2))\\
    {\hfpred{p_1}{c}{f_{r1}}{v^*_1} {\sep}
    \heap_1{\wedge}\pure_1{\wedge}\tau_1'}
    \quad \vdash^{\heap'}_{V}
    {\heap_2{\wedge}\pure_2{\wedge}\tau_2{\wedge}\rho} \,
    {\sep} \, {\D}
	\end{array}}
  \]
  \\
  \[
	  \infer[\begin{array}{l} \mathsf{RIGHT}\textrm{-} \\ \mathsf{SPLIT}\textrm{-}
	  \\
	  \mathsf{MATCH}\textrm{ (c)} \end{array}] {\entailVV{\hfpred{p_1}{c}{f_1}{v^*_1}{\sep}\heap_1{\wedge}\pure_1{\wedge}\tau_1}
	{(\hfpred{p_2}{c}{f_2}{v^*_2}{\sep}\heap_2{\wedge}\pure_2{\wedge}\tau_2)}{\D} }
	{\begin{array}{c}
	\mathsf{fst}(\fxpure(\hfpred{p_1}{c}{f_1}{v^*_1}{\sep}\heap_1,
	\tau_1)){\wedge}\pure_1 ~~ \implies ~~ p_1{=}p_2 \\
	V'= \code{if} ~f_2 \in V ~\code{then} ~V \cup \{f_{c2},f_{r2}\} ~\code{else} ~V\\
	\tau_1'= \code{if} ~f_2 \in V ~\code{then} ~\tau_1 ~\code{else} ~(\tau_1
	{\wedge}f_{c2}\oplus f_{r2} = f_2) \\
	\tau_2'= \code{if} ~f_2 \in V ~\code{then} ~(\tau_2 {\wedge}f_{c2}\oplus f_{r2} = f_2) ~\code{else}
	~\tau_2 \\
	\heap' = \heap{\sep}\hfpred{p_1}{c}{f_1}{v^*_1} \\
	\rho = f_1{=}f_{c2}{\wedge}(\bigwedge_i(v^i_1=v^i_2))\\
	{\heap_1{\wedge}\pure_1{\wedge}\tau_1'}
    \vdash^{\heap'}_{V'}
	{~~ \hfpred{p_2}{c}{f_{r2}}{v^*_2} {\sep}
	\heap_2{\wedge}\pure_2{\wedge}\tau_2'\wedge\rho\sep \D}
	\end{array}}
  \]
  Because the search can be computationally expensive, we have devised an aggressive pruning technique.  
  We try to determine to what extent the fractional constraints restrict the fractional variables. It may be that (a) $f_1=f_2$, in which case only
  $\mathsf{FULL}\textrm{-}\mathsf{MATCH}$ is feasible, or (b) $f_1$ is included in $f_2$, in which case
  $\mathsf{RIGHT}\textrm{-}\mathsf{SPLIT}\textrm{-}\mathsf{MATCH}$ is feasible, or (c) $f_1$ includes $f_2$,
  in which case only\\
   $\mathsf{LEFT}\textrm{-}\mathsf{SPLIT}\textrm{-}\mathsf{MATCH}$ is feasible.

\end{iteMize}

\subsection{Proving barrier soundness}
\label{sec:toolbarsound}

The fractional share solver and enhancements to SLEEK's entailment procedures discussed above help
with any program logic that needs fractional shares (\emph{e.g.}, concurrent separation logic with locks,
sequential separation logic with read-only data).  In contrast, our other enhancements are
specific to the logic for Pthreads-style barriers.  Our initial goal is to automatically check
the consistency of barrier definitions---that is, whether a barrier definition meets the side conditions
presented in \S\ref{sec:bardefs}.  The first step is to describe a barrier diagram
to SLEEK.

Although the barrier diagrams presented in \S\ref{sec:bardefs} are intuitive and concise, programs
need a more textual representation.  Barrier diagrams describe the possible transitions a
barrier state can make and the specifications associated with those transitions.  In a sense, a
barrier definition can be viewed as a disjunctive predicate definition where the body is a
disjunction of possible transitions.

SLEEK already contains user-defined predicates so it is easy to introduce the ``is a barrier''
predicate $barrier(bn,v_f,s)$ as required by the barrier logic, with a slight change to notation to
accommodate the syntax presented in Figure \ref{SLEEKlang} to $bn^{v_f}{\langle}s{\rangle}$.

We extended SLEEK's language to accept barrier diagrams in the form:
\[
\begin{array}{lcl}
\mathsf{bdef} &::=& \code {barrier} ~(\mathit{b\_name}~,~ \mathit{thread\_cnt} ~,~ v^* ~,~ \mathsf{transition}^*)\\
\mathsf{transition} &::=& (\mathit{from\_state} ~,~\mathit{to\_state} ~,~ \mathsf{pre}\textrm{-}\mathsf{post}\textrm{-}\mathsf{spec}^*) \\
\mathsf{pre}\textrm{-}\mathsf{post}\textrm{-}\mathsf{spec} &::=& (\Phi_{pre} ~,~ \Phi_{post}) \\
\end{array}
\]
SLEEK can now automatically check the well-formedness conditions on the barrier definitions as follows:
\begin{iteMize}{$\bullet$}
  \item All transitions must have exactly $\mathit{thread\_cnt}$ specifications, one for each
      thread
  \item For each transition, let $\mathit{from}$ and $\mathit{to}$ be the state labels, then:
  	\begin{iteMize}{$-$}
  	  \item  for each specification $(\Phi_{\mathit{pre}} ~,~ \Phi_{\mathit{post}})$
  	  \begin{enumerate}[(1)]
  	    \item  $\Phi_{\mathit{pre}}$ contains a fraction of the barrier in state $\mathit{from}$: \\
  	    $\Phi_{\mathit{pre}}\vdash \hfpred {\mathsf{self}}{bn}{v_f}{\mathit{from}} ~
  \sep ~ \Delta$
  	    \item  $\Phi_{\mathit{post}}$ contains a fraction of the barrier in state $\mathit{to}$:\\
  	     $\Phi_{\mathit{pre}}\vdash  \hfpred {\mathsf{self}}{bn}{v_f}{\mathit{to}} ~
  \sep ~ \Delta$
  	    \item  $\Phi_{\mathit{pre}}\sep \Phi_{\mathit{pre}} \vdash \mathsf{False}$

  \qquad The soundness proof assumes that each precondition $P$ is precise.
  Unfortunately, precision is not very easy to verify automatically.  As indicated in
  footnote 7, we believe that the logic will be sound if we can assume the (strictly)
  weaker property ``token'': $P \star P \vdash \mathsf{False}$ instead of precision. At
  this stage, our prototype extension to SLEEK verifies that preconditions are tokens
  rather than that they are precise.  We are in the process of attempting to update our
  soundness proof to require that preconditions be tokens rather than precise; if we
  are unable to do so then one solution would be for SLEEK to output a Coq file stating
  lemmas regarding the precision of each precondition.  Users would then be required to
  prove these lemmas manually to be sure that their barrier definitions were sound.  In
  our example barrier, the Coq proofs of precision were only a small part of the 2,700
  total lines of Coq script, so the savings from using SLEEK to verify the soundness of
  a barrier definition should still be quite substantial.  Another choice would be to
  devise a heuristic algorithm for determining precision; we suspect that such an
  algorithm could handle the examples from this paper.

	  \end{enumerate}
	  \item the star of all the preconditions contains the full barrier (recall that the
    entailment $\vdash$ check in SLEEK can produce a residue) \\
	  	$\sep_{i=1}^{i=\mathit{thread\_cnt}}\Phi_{\mathit{pre}}^i \vdash \hfpred
{\mathsf{self}}{bn}{\fullp}{\mathit{from}} ~ \sep ~ \Delta $
	  \item the star of all the postconditions contains the full barrier\\
	  	$\sep_{i=1}^{i=\mathit{thread\_cnt}}\Phi_{\mathit{post}}^i \vdash \hfpred
{\mathsf{self}}{bn}{\fullp}{\mathit{to}} ~ \sep ~ \Delta$
	  \item the star of all the preconditions equals the star of all postconditions modulo
the barrier state change for a transition.  We check this constraint by carving the full
barrier out of the total heap using the residues $\D_\mathit{pre}$ and $\D_\mathit{post}$
of the entailments given in the previous constraints.  $\D_\mathit{pre}$ and $\D_\mathit{post}$ are
then tested for equality by requiring bi-entailment with empty residue.  That is, given \\
 $~~~~ \sep_{i=1}^{i=\mathit{thread\_cnt}}\Phi_{\mathit{pre}}^i \vdash \hfpred {\mathsf{self}}{bn}{\fullp}{\mathit{from}}\sep \D_{\mathit{pre}}$ \\
and \\
     $~~~~ \sep_{i=1}^{i=\mathit{thread\_cnt}}\Phi_{\mathit{post}}^i \vdash \hfpred{\mathsf{self}}{bn}{\fullp}{\mathit{to}}\sep \D_{\mathit{post}}$, \\
we check \\
 $~~~~ \D_{\mathit{pre}}\vdash \D_{\mathit{post}} ~~ \textrm{and} ~~
  \D_{\mathit{post}}\vdash \D_{\mathit{pre}}$ \textbf{with empty residues}.
  	  \item For states with more than one successor, we check mutual exclusion for the
  preconditions as required by \S\ref{sec:bardefs} by verifying that for any two
  preconditions of two distinct transitions must entail $\mathsf{False}$.  This check was
   extremely tedious to do for the example barrier in Coq but SLEEK can do it easily.
	\end{iteMize}
\end{iteMize}
Once SLEEK has verified each of the above conditions, the barrier definition is
well-formed according to the constraints described in \S\ref{sec:bardefs} (modulo precision).

\subsection{Extension to program verification}
Integrating our Hoare rule for barriers into HIP was the easiest part of adding our program logic
to HIP/SLEEK.  Following the concept of structured specifications \cite{FM11:Gherghina}, we
transform our barrier diagrams into disjunctions of the form
\[
bn ~::=~ \bigvee (\code{requires}~ \Phi_{pre} ~~~~ \code{ensures} ~ \Phi_{post}),
\]
where the disjunction spans all specifications in all transitions in the barrier definition.
Verification of barrier calls trivially reduces to an entailment check of the disjunction.

\subsection {Tool performance outline}
We have developed a small set of benchmarks for our HIP/SLEEK with barriers prototype.  Our SLEEK
tests divide into two categories: entailment checks for separation logic formulae containing
fractional permissions and checking barrier consistency checks as in \S\ref{sec:toolbarsound}.
Individual entailment checks are quite speedy and our benchmark covers a number of interesting
cases (\emph{e.g.}, inductively defined predicates).  Barrier consistency checks take more time but
the performance is more than adequate:
\[
\begin{tabular}{|l|c|c|c|}
\hline
~~ Sleek Examples ~~ & Test details & ~~ Entailments ~~ & ~~ Time (s) ~~ \\
\hline
~~\texttt{fractions.slk} & ~~ fractional entailments ~~ & 54 & 0.08 \\
~~\texttt{barrier.slk}   & 6 barrier definitions & 279 & 2.3 \\
\hline
\end{tabular}
\]
One of the barrier definitions in \texttt{barrier.slk} is the example barrier given in
Figure~\ref{fig:ex_code}.  It took \textbf{2,700 (highly tedious) lines of code} and 48 seconds of
verification time (Figure \ref{tbl:proof}) to convince Coq that the example barrier definition met
the soundness requirements\footnote{Techniques such as those developed by Braibant \emph{et al.} \cite{bp:cpp11:aac},
Nanevski \emph{et al.} \cite{nanevski:popl10}, and Gonthier \emph{et al.} \cite{gonthier:icfp11}
can probably eliminate some (but not all) of the tedium of reasoning about the associativity and
commutativity of $\sep$. Unfortunately, proofs of mutual exclusion for
barrier transitions seem less tractable.  An alternative approach would be to use a
separation logic entailment procedure implemented in Coq such as the one recently described by
Appel \cite{appel:cpp11}.}. SLEEK verifies this example barrier definition and analyzes five others
(some sound, others not) in 2.3 seconds \textbf{without any interaction from the user}\footnote{As
explained in \S\ref{sec:toolbarsound}, SLEEK verifies properties that are slightly different from those
verified in Coq.}.

We have also benchmarked HIP with a slightly modified variant of the example from \S
\ref{sec:example}.  We made two modifications due to certain existing weaknesses in the HIP/SLEEK
toolchain.  First, we substituted recursive functions for loops due to the convenience of
specification of recursive functions in HIP; and second, we changed the way $x_1$, $x_2$, $y_1$,
and $y_2$ are modified in lines 6 and 10 to enable the numerical decision procedures (\emph{i.e.},
Omega) to discharge the associated obligations.  Both of these changes are orthogonal to
our logic for barriers: for example, a more powerful decision procedure for numerical equalities
would allow us to return to the original program.

We verified our modified code against three specifications.  In \texttt{barrier-paper.ss}, we
verify a trivial correctness property for the exact barrier definition from
\ref{sec:example}---\emph{i.e.}, we verify the postcondition of \texttt{True}, meaning that the
program does not get stuck.  We also verified two more complex postconditions by using two more
finely-grained barrier definitions: in \texttt{barrier-weak.ss}, we verified the relationship
between $x_1$, $x_2$, and $n$; finally, in \texttt{barrier-strong.ss} we verified the precise value
in $x_1$ after the loop terminates (\emph{i.e.}, $x_1 = 59$).  The code and specification for
\texttt{barrier-strong.ss} is given in Appendix \ref{sec:appendix}.  We recorded the following
timings from HIP:
\[
\begin{tabular}{|l|c|c|c|}
\hline
~ File & ~ Postcondition ~ & ~ LOC (code + specification) ~ & ~ Time(s) ~\\
\hline
~\texttt{barrier-paper.ss} & \texttt{True} & 73 & 2.55 \\
~\texttt{barrier-weak.ss} & lax bounds & 73 & 2.91 \\
~\texttt{barrier-strong.ss} ~ & exact bounds & 73 & 3.04 \\
\hline
\end{tabular}
\]
As expected, the tighter bounds require more verification time; however, the differences are
relatively small because most of the work is dealing with the heap constraints as opposed to
the pure constraints.  Part of the time for each example is spent verifying the correctness of the
included barrier definition; all three barrier definitions from the HIP examples were also included
in the \texttt{barrier.slk} benchmark.

HIP verification times are decent, but barrier calls are fairly
computationally expensive to verify due to the need to check multiple entailments.  We believe that
performance can be further improved by adding optimizations to SLEEK in the style of
\cite{DBLP:conf/cav/ChinGVLCQ11}.  Since barrier calls are fairly rare in actual code, we believe
that the performance of HIP/SLEEK on larger examples will be acceptable.



\section {Limitations and Future Work}

We can extend the logic by making the barriers first-class (i.e., dynamic barrier
creation/ destruction).  In the present work we thought we could simplify the proofs by having
statically declared barriers in the style of O'Hearn \cite{OHearn05resconc}.  This turned out to be
somewhat of a mistake, at least as far as the soundness proof went: since we were forced to track
the barrier states (and partial shares) explicitly in the Hoare logic, we estimate that 90\% of the
work required to make the barriers first-class has already been done in the present work; moreover,
a further 8\% (the intrinsic contravariant circularity) would be easy to handle via indirection
theory \cite{hobor10:popl}. With perfect foresight (or if it were trivial to restart a large
mechanized proof), we would have certainly made the barriers first-class.  Our SLEEK prototype does
support first-class barriers using the barrier creation rule we expect to be true.


We suspect that our SLEEK prototype could be improved in numerous ways.  For example, our decision
procedure for share formulae is quite incomplete\footnote{For example, we cannot verify $\forall
\pi_1,\pi_2,\pi_3.~~~ \pi_1 \oplus \pi_2 = \pi_3 ~~ \vdash ~~ \pi_1 \oplus \pi_2 = \pi_3 $.} and we
believe that several performance enhancements to SLEEK would speed up the consistency checks.
Finally, we need to resolve the precision/token issue.

We also do not address the tricky problem of barrier definition inference. \hide{  We would like to
investigate inferring the barrier definition automatically. verifying program text containing
barrier calls; one place to begin is constructing a verifier for programs that use OpenMP
\cite{OpenMPBook}. }
\section{Related Work}

Calcagno \emph{et al}. proposed separation algebras as models of separation logic
\cite{calcagno2007}; fractional permissions were discussed by Bornat \emph{et al}.
\cite{bornat05:popl}.  In our work we use the share model and separation algebra development of
Dockins \emph{et al}. \cite{dockins09:sa,MSL}.

O'Hearn's concurrent separation logic focused on programs that used critical regions
\cite{OHearn05resconc,Brookes04}; subsequent work by Hobor \emph{et al}. and Gotsman \emph{et al}.
added first-class locks and threads \cite{HoborAN08,GotsmanBCRS07,Hobor:Thesis}. Our basic
soundness techniques (unerased semantics tracks resource accounting; oracle semantics isolates
sequential and concurrent reasoning from each other; etc.) follow Hobor \emph{et al}. Recently both
Villard \emph{et al}. and Bell \emph{et al}. extended concurrent separation logic to channels \cite
{BellAW10,villard:aplas09}.  The work on channels is similar to ours in that both Bell and Villard
track additional dynamic state in the logic and soundness proof.  Bell tracks communication
histories while Villard tracks the state of a finite state automaton associated with each
communication channel.  Of all of the previous soundness results, only Hobor \emph{et al}. had a
machine-checked soundness proof, albeit an incomplete one.

An interesting question is whether is it possible to reason about barriers in a setting with locks
or channels.  The question has both an operational and a logical flavor.  Speaking operationally,
in a practical sense the answer is no: for performance reasons barriers are not implemented with
channels or locks.  If we ignore performance, however, it \textbf{is} possible to implement
barriers with channels or locks\footnote{Indeed, it is possible to implement channels and locks in
terms of each other.}.  The logical part of the question then becomes, are the program logics
defined by O'Hearn, Hobor, Gotsman, Villard, or Bell (including their coauthors) strong enough to
reason about the (implementation of) barriers in the style of the logic we have presented? As far
as we can tell each previous solution is missing at least one required feature, so in a strict
sense, the answer here is again no.

For illustration we examine what seems to be the closest solution to ours: the copyless message
passing channels of Villard \emph{et al.}  Operationally speaking, the best way to implement
barriers seems to be by adding a central authority that maintains a channel with each thread using
a barrier. When a thread hits a barrier, it sends ``waiting'' to the central authority, and then
waits until it receives ``proceed''.  In turn, the central authority waits for a ``waiting''
message from each thread, and then sends each of them a ``proceed'' message. Fortunately Villard
allows the central authority to wait on multiple channels simultaneously.

The question then becomes a logical one.  Although it should not pose any fundamental difficulty,
their logic would first need to be enhanced with fractional permissions; in fact we believe that
Villard's Heap-Hop tool already uses the same fractional permission model (by Dockins \emph{et
al.}) that we do\footnote{To be precise, Heap-Hop uses the code extracted from the fractional
permission Coq proof development by Dockins \emph{et al.}}.  Since Villard uses automata to track
state, we think it probable, but not certain, that our barrier state machines can be encoded as a
series of his channel state machines.

There are some problems to solve.  Villard requires certain side conditions on his channels; we
require other kinds of side conditions on our barriers; these conditions do not seem fully
compatible\footnote{For example, Villard requires determinacy whereas we do not; he would also
require that the postconditions of barriers be precise whereas we do not; etc.}.  Assuming that we
can weaken/strengthen conditions appropriately, we reach a second problem with the side conditions:
some of our side conditions (\emph{e.g.}, mutual exclusion) are restrictions on the shape of the
entire diagram; in Villard's setting the barrier state diagram has been partitioned into numerous
separate channel state machines.  Verifying our side conditions seems to require verification of
the relationships that these channel state machines have to each other; the exact process is
unclear.

Once the matter of side conditions is settled, there remains the issue of verifying the individual
threads and the central authority.  Villard's logic seems to have all that is required for the
individual threads; the question is how difficult it would be to verify the central authority.
Here we are less sure but suspect that with enough ghost state/instructions it can be done.

There remains a question as to whether it is a good idea to reason about barriers via channels (or
locks). We suspect that it is not a good idea, even ignoring the fact that actual implementations
of barriers do not use channels.  The main problem seems to be a loss of intuition: by distributing
the barrier state machine across numerous channel state machines and the inclusion of necessary
ghost state, it becomes much harder to see what is going on.  We believe that one of the major
contributions of our work is that our barrier rule is extremely simple; with a quick reference to
the barrier state diagram it is easy to determine what is going on.  There is a secondary problem:
we believe that our barrier rule will look and behave essentially the same way in a setting with
first-class barriers in which it is possible to define functions that are polymorphic over the
barrier diagram; even assuming a channel logic enriched in a similar way, the verification of a
polymorphic central authority seems potentially formidable.

One interesting question is how our barrier rule would interact with the rules of other flavors of
concurrent separation logic (\emph{e.g.}, with locks or channels).  We believe that the answer is
yes, at least in the context of a logic of partial correctness\footnote{Of course, the more
concurrency primitives a programmer has, the easier it is to get into a deadlock.  We hypothesize
that concurrent program logics of total correctness may not be as compositional as concurrent
program logics of partial correctness.}, as long as the primitives used remain strongly
synchronizing (\emph{i.e.}, coarse-grained).  It is not clear how our barrier rule might interact
with the kind of fine-grained concurrency that is the subject of Vafeiadis and Parkinson
\cite{vafeiadis:concur07}, Dodds \emph{et al.} \cite{dodds:esop09}, or Dinsdale-Young \emph{et al.}
\cite{dinsdale-young:ecoop10}.  We believe that our barrier rule is sound on a machine with weak
memory as long as all of the concurrency is strongly synchronized.



Finally, work on concurrent program analysis is in the early stages; Gotsman \emph{et al}.,
Calcagno \emph{et al}., and Villard \emph{et al}. give techniques that cover some use cases
involving locks and channels but much remains to be done
\cite{Gotsman:SAS06,Calcagno:APLAS09,villard:tacas10}.

\paragraph{Connection to a result by Jacobs and Piessens.}  We recently learned that
Jacobs and Piessens have an impressive result on modular fine-grained concurrency
\cite{jacobs:popl11}.  Jacobs was able to reason about our example program using his VeriFast tool
by designing an implementation of barriers using locks and reducing our barrier diagram to a large
disjunction for a resource invariant.

However, VeriFast has some disadvantages compared to the HIP/SLEEK approach we presented. First,
HIP/SLEEK required far less input from the user.  In the case of our 30-line example program, more
than 600 lines of annotation were required in VeriFast, not including the code/annotiations for the
barrier implementation itself.  HIP/SLEEK were able to verify the same program with approximately
30 lines of annotation (mostly the barrier definition).  Second, it was harder to gain insight into
the program from the disjunction-form of the invariant; in contrast we find our barrier diagrams
straightforward to understand.  Finally, it is unclear to us whether the reduction is always
possible or whether it was only enabled by the relative simplicity of our example program. That
said, Jacobs and Piessens have the only logic and tool proven to be able to reason about barriers
as derived from a more general mechanism.

\section{Conclusion}

We have designed and proved sound a program logic for Pthreads-style barriers.  Our development
includes a formal design for barrier definitions and a series of soundness conditions to verify
that a particular barrier can be used safely.  Our Hoare rules can verify threads independently,
enabling a thread-modular approach. Our soundness proof defines an operational semantics that
explicitly tracks permission accounting during barrier calls and is machine-checked in Coq. We have
modified the verification toolset HIP/SLEEK to use our logic to verify concurrent programs that use
barriers.

~

Our soundness results are machine-checked in Coq and are available at:
		\[{\texttt{www.comp.nus.edu.sg/{$\sim$}hobor/barrier}}\]

Our prototype HIP/SLEEK verification tool is available at:
  		\[{\texttt{www.comp.nus.edu.sg/{$\sim$}cristian/projects/barriers/tool.html}}\]

\paragraph{Acknowledgements.}  We thank Christian Bienia for showcasing numerous example
programs containing barriers, Christopher Chak for help on an early version of this work, Jules
Villard for useful comments in general and in particular on the relation of our logic to the logic
of his Heap-Hop tool, and Bart Jacobs for discovering how to verify our example program in his
VeriFast tool.


\bibliography{barrier}
\bibliographystyle{plain}
\newpage
\appendix
\section {Example from \S\ref{sec:example} revisited} \label{sec:appendix}

Below we present a slight variation on the example from section \S\ref{sec:example} that we
verified with our HIP/SLEEK toolchain.  In this example, we specify exact postconditions.  Starting
the execution with $x_1=x_2=1$ will lead to $x_1=x_2=59$.  The example is expressed in the
HIP/SLEEK input language (where [L] and [R] respectively denote the left and right half of the full
share).  It makes use of recursive functions instead of while loops, but this is only for aesthetic
reasons.
\\
\\
\begin{verbatim}
data cl {int val;}

barrier bn, 2,x1 x2 y1 y2 i, 
/* bn-barrier name, 2-thread count, x1..i shared heap                           */
/* the list of shared variables denotes the arguments of the barrier definition */
/* however for technical reasons we found it easier to list the variables here  */
/* these could be infered with some additional work                             */   
[(0,1,   // transition description, start/end state
  [ requires
        x1::cl@[L]<A1>*x2::cl@[L]<B1>* y1::cl@[L]<C1>*y2::cl@[L]<D1>*
        i::cl@[L]<T1>*self::bn@[L]<0>
    ensures
        x1::cl@[L]<A1>*x2::cl@[L]<B1>* y1::cl<C1>*i::cl@[L]<T1>*
        self::bn@[L]<1> & T1 < 30;,                             // one pre-post
    requires
        x1::cl@[R]<A2>*x2::cl@[R]<B2>*y1::cl@[R]<C2>*y2::cl@[R]<D2>*
        i::cl@[R]<T2>*self::bn@[R]<0>
    ensures
        x1::cl@[R]<A2>*x2::cl@[R]<B2>*y2::cl<D2>*i::cl@[R]<T2>*
        self::bn@[R]<1> & T2 < 30;]),
	
 (1,2,[
    requires
        x1::cl@[L]<A>*x2::cl@[L]<A>*y1::cl<C>* i::cl@[L]<T>* self::bn@[L]<1>&
        T<30 & A=2*T-1 & C = 3*A+2
    ensures
        x1::cl<A>*y1::cl@[L]<C>*y2::cl@[L]<D>*i::cl<T>*self::bn@[L]<2>&
        T<30 & A=2*T-1 & D=2*A & C = 3*A+2;,
    requires
        x1::cl@[R]<A>*x2::cl@[R]<A>*y2::cl<D>*i::cl@[R]<T>*self::bn@[R]<1>&
        T<30 & D=2*A & A=2*T-1
    ensures
        x2::cl<A>*y1::cl@[R]<C>*y2::cl@[R]<D>* self::bn@[R]<2> &
        D=2*A & C = 3*A+2 & A=2*T-1;]),

 (2,1,[
    requires
        x2::cl<B>*y1::cl@[R]<C>*y2::cl@[R]<D>* self::bn@[R]<2> & B=C-D
    ensures
        x1::cl@[R]<A>*x2::cl@[R]<B>*y2::cl<D>*i::cl@[R]<T>*self::bn@[R]<1>&
        A=C-D & A=B & A=2*T-1 & T <= 30;,
    requires
        x1::cl<A>*y1::cl@[L]<C>*y2::cl@[L]<D>*i::cl<T>*self::bn@[L]<2>&
        A=C-D & A=2*T-1 & T <= 30
    ensures
        x1::cl@[L]<A>*x2::cl@[L]<B>*y1::cl<C>*i::cl@[L]<T>*self::bn@[L]<1>&
        A=C-D & A=B & A=2*T-1 & T <= 30;]) ,

 (1,3,[
    requires
        x1::cl@[L]<A>*x2::cl@[L]<B>*i::cl@[L]<T>*self::bn@[L]<1>& T=30
    ensures
        x1::cl@[L]<A>*x2::cl@[L]<B>*i::cl<T>*self::bn@[L]<3> & T=30;,
    requires
        x1::cl@[R]<A>*x2::cl@[R]<B>*i::cl@[R]<T>*self::bn@[R]<1>& T=30
    ensures
        x1::cl@[R]<A>*x2::cl@[R]<B> *self::bn@[R]<3>;])];

// end barrier definition, begin code

void th1 (cl x1, cl x2, cl y1, cl y2, cl i, bn b)
    requires
        x1::cl@[L]<1>*x2::cl@[L]<1>*y1::cl@[L]<_>*
        y2::cl@[L]<_>*i::cl@[L]<1>*b::bn@[L]<0>
    ensures
        x1::cl@[L]<v>*x2::cl@[L]<v>*b::bn@[L]<3>& v=59;
{                                 // stage 0
  barrier b;                      // stage 0->1
  th1_loop (x1,x2,y1,y2,i,b);
}

void th1_loop(cl x1, cl x2, cl y1, cl y2, cl i, bn b)
    requires
        x1::cl@[L]<v>*x2::cl@[L]<v>*y1::cl<_>*i::cl@[L]<a>*
        b::bn@[L]<1> & v=2*a -1 & a <= 30
    ensures
        x1::cl@[L]<v1>*x2::cl@[L]<v1>*b::bn@[L]<3>& v1=59;
{
  if (i.val<30)
  {                               // stage 1
    y1.val = x1.val + 2*x2.val+2;
    barrier b;                    // stage 1->2
    x1.val = y1.val - y2.val;
    i.val= i.val+1;
    barrier b;                    // stage 2->1
    th1_loop (x1,x2,y1,y2,i,b);
  }
  else barrier b;                 // stage 1->3
}

void th2 (cl x1, cl x2, cl y1, cl y2, cl i, bn b)
    requires
        x1::cl@[R]<1>*x2::cl@[R]<1>*y1::cl@[R]<_>*y2::cl@[R]<_>*
        i::cl@[R]<1>*b::bn@[R]<0>
    ensures
        x1::cl@[R]<v>*x2::cl@[R]<v>*b::bn@[R]<3>& v=59;
{                                 // stage 0
  barrier b;                      // stage 0->1
  th2_loop (x1,x2,y1,y2,i,b);
}

void th2_loop(cl x1, cl x2, cl y1, cl y2, cl i, bn b)
    requires
        x1::cl@[R]<v>*x2::cl@[R]<v>*y2::cl<_>*i::cl@[R]<a>*
        b::bn@[R]<1> & v=2*a -1 & a <= 30
    ensures
        x1::cl@[R]<v1>*x2::cl@[R]<v1>*b::bn@[R]<3>& v1=59;
{
  if (i.val<30)
  {                               // stage 1
    y2.val = x1.val + x2.val;
    barrier b;                    // stage 1->2
    x2.val = y1.val - y2.val;
    barrier b;                    // stage 2->1
    th2_loop (x1,x2,y1,y2,i,b);
  }
  else barrier b;                 // stage 1->3
}
\end{verbatim}

\end{document}